\documentclass[5p]{elsarticle} 

\usepackage{mathtools}
\usepackage{enumitem}

\usepackage[table,xcdraw]{xcolor}
\journal{Environmental Research Letters}

\usepackage{placeins}

\usepackage{natbib}
\usepackage{eurosym}
\usepackage{threeparttable} 
\usepackage{subcaption}
\usepackage{url}
\usepackage{xurl} 
\usepackage[colorlinks=true, citecolor=blue, linkcolor=blue, filecolor=blue,urlcolor=blue]{hyperref}
\usepackage[acronym, automake, nonumberlist]{glossaries-extra}
\usepackage{svg}

\usepackage{lineno}
\modulolinenumbers[5]

\usepackage{lineno,hyperref}
\modulolinenumbers[1]
\usepackage{amsmath}
\usepackage{siunitx}
\usepackage{eurosym}
\biboptions{numbers,sort&compress}
\usepackage[europeanresistors,americaninductors]{circuitikz}
\usepackage{adjustbox}
\usepackage{xspace}
\usepackage{caption}
\usepackage{booktabs}
\usepackage{tabularx}
\usepackage{threeparttable}
\usepackage{multicol, multirow}
\usepackage{float}
\usepackage{graphicx,dblfloatfix}
\usepackage{csvsimple}
\usepackage{amssymb}
\usepackage{pifont}
\usepackage{tikzsymbols}
\usepackage{textcomp}

\newcolumntype{L}{>{\begin{math}}l<{\end{math}}}%
\newcolumntype{C}{>{\begin{math}}c<{\end{math}}}%

\newcommand{\mwhh}{~\officialeuro /MWh$_{\text{H}_2}$}

\newcommand{\eurt}{~\officialeuro/t$_{\text{CO}_2}$}

\newcommand{\ce}[1]{{#1}$^\text{o}$C}

\makeglossaries
\newacronym{icct}{ICCT}{International Council on Clean Transportation}
\newacronym{tes}{TES}{Thermal energy storage (in form of water tanks)}
\newacronym{chp}{CHP}{Combined heat and power plant}
\newacronym{cop}{COP}{Coefficient of Performance}
\newacronym{bdew}{BDEW}{Bundesverband der Energie- und Wasserwirtschaft}
\newacronym{PV}{PV}{Solar photovoltaics}
\newacronym{ghg}{GHG}{Greenhouse gas}
\newacronym{EU}{EU}{European Union}
\newacronym{ev}{EV}{Electric Vehicle}
\newacronym{bev}{BEV}{Battery Electric Vehicle}
\newacronym{dea}{DEA}{Danish Energy Agency}
\newacronym{lng}{LNG}{Liquified Natural Gas}
\newacronym{cng}{CNG}{Compressed Natural Gas}
\newacronym{helmeth}{HELMETH}{Integrated High-Temperature Electrolysis and Methanation for Effective Power to Gas Conversion}
\newacronym{jrc}{JRC}{Joint Research Center}
\newacronym{idees}{IDEES}{Integrated Database of the European Energy System}
\newacronym{iam}{IAM}{Integrated Assessment Model}
\newacronym{ecmwf}{ECMWF}{European Centre for Medium-Range Weather Forecasts}
\newacronym{ocgt}{OCGT}{Open cyclic gas turbine}
\newacronym{v2g}{V2G}{Vehicle to Grid}
\newacronym{pypsa}{PyPSA}{Python for Power System Analysis}
\newacronym{tyndp}{TYNDP}{Ten-year network development plan}
\newacronym{entso-e}{ENTSO-E}{European network for transmission system operators electricity}
\newacronym{nuts}{NUTS}{Nomenclature of Territorial Units for Statistics}
\newacronym{covid}{COVID-19}{Coronavirus disease 2019}
\newacronym{entsog}{ENTSO-G}{European Network of Transmission System Operators for Gas}
\newacronym{dsm}{DSM}{Demand side management}
\newacronym{hvdc}{HVDC}{High voltage direct current}
\newacronym{phs}{PHS}{Pumped hydro storage}
\newacronym{dac}{DAC}{Direct air capture}
\newacronym{sng}{SNG}{Synthetic natural gas}
\newacronym{ise}{Fraunhofer ISE}{Fraunhofer Institute for Solar Energy Systems}
\newacronym{diw}{DIW}{German Institute for Economic Research (Deutsches Institut für Wirtschaftsforschung)}
\newacronym{SMR}{SMR}{Steam methane reforming}
\newacronym{fom}{FOM}{Fixed operation and maintenance costs}
\newacronym{vom}{VOM}{Variable operation and maintenance costs}
\newacronym{ppa}{PPA}{Power Purchase Agreement}
\newacronym{ptc}{PTC}{Production Tax Credit}
\newacronym{lohc}{LOHC}{Liquid organic hydrogen carriers}
\newacronym{res}{RES}{Renewable energy sources}
\newacronym{necp}{NECP}{National energy climate plan}
\newacronym{ice}{ICE}{Internal Combustion Engine}
\newacronym{fcev}{FCEV}{Fuel Cell Electric Vehicle}
\newacronym{DA}{DA}{Delegated Act for the regulation on Union methodology for renewable fuels of non-biological origin}
\newacronym{ets}{ETS}{Emissions Trading System}
\newacronym{capex}{CAPEX}{Capital expenditures}
\newacronym{cfd}{CfD}{Contract for Difference}
\newacronym{lulucf}{LULUCF}{land use, land-use change, and forestry}
\newacronym{bast}{BASt}{German Federal Highway Research Institute}
\begin{document}

\begin{frontmatter}

\title{Shifting burdens: How delayed decarbonisation of road transport affects other sectoral emission reductions}

\author[tubaddress]{Elisabeth Zeyen}
\author[aarhusaddress]{Sina Kalweit}
\author[aarhusaddress]{Marta Victoria\textsuperscript{c,}}
\author[tubaddress]{Tom Brown}
\address[tubaddress]{Department of Digital Transformation in Energy Systems, Faculty of Process Engineering, TU Berlin, Einsteinufer 25 (TA 8),Berlin, 10587, Berlin,Germany}
\address[aarhusaddress]{Department of Mechanical and Production Engineering and iCLIMATE Interdisciplinary Centre for Climate Change, Aarhus University, 8000, Aarhus, Denmark}
\address[dtuaddress]{Department of Wind and Energy Systems, Technical University of Denmark, Elektrovej 325, 2800, Lyngby, Denmark}

\begin{abstract}
In 2022, fuel combustion in road transport accounted for approximately 21\% (760 million tonnes) of CO$_2$ emissions in the European Union (EU). Road transport is the only sector with rising emissions, with an increase of 24\% compared to 1990. The EU initially aimed to ban new CO$_2$-emitting cars by 2030 but has since delayed this target to 2035, underscoring the ongoing challenges in the push for rapid decarbonisation. The pace of decarbonisation in this sector will either ease or intensify the pressure on other sectors to stay within the EU’s carbon budget. This paper explores the effects of speeding up or slowing down the transition in road transport. We reveal that a slower decarbonisation path not only drives up system costs by 126~billion \officialeuro/a (6\%) but also demands more than a doubling of the CO$_2$ price from 137 to 290\eurt in 2030 to trigger decarbonisation in other sectors. On the flip side, accelerating the shift to cleaner transport proves to be the most cost-effective strategy, giving room for more gradual changes in the heating and industrial sectors, while reducing the reliance on carbon removal in later years. Earlier mandates than currently envisaged by the EU can avoid stranded assets and save up to 43~billion \officialeuro/a compared to current policies.

\end{abstract}


\end{frontmatter}


\section*{Highlights}

\begin{itemize}
\item Examines trade-offs between the speed of road transport decarbonisation and CO$_2$ reduction burdens in other sectors.
\item Electrification is the cost-optimal decarbonisation strategy for both light- and heavy-duty transport, even under significant vehicle cost variations.
\item Faster road transition reduces costs, allows for slower build out of heat pumps and green hydrogen production and reduces reliance on \gls{dac}
\item Mandates for zero-emission vehicles by 2030 (light-duty) and 2035 (heavy-duty) save costs and prevent stranded assets.
\item Road transport demand reductions lower long-term costs and renewable energy capacity requirements.
\end{itemize}

\section{Introduction}
Over the past three decades, Europe has reduced its greenhouse gas emissions, achieving a 27\% reduction between 1990 and 2021. However, this overall reduction masks a stark disparity between sectors: while most manage to reduce emissions, road transport emissions increased by 21\% during the same period \cite{EEA_National_Emissions_2024}. This increase highlights the persistent challenge of decarbonising this sector and is primarily attributed to the continuously growing demand, driven by rising vehicle-kilometres travelled and the trend toward larger and more powerful cars, despite advancements in vehicle efficiency.

Currently road transport predominantly relies on fossil fuel combustion, making it one of the largest contributors to greenhouse gas emissions. Next to increasing the use of public transport and reducing demand, the primary strategies to decarbonise the sector involves transitioning from fossil-fueled internal combustion engines (\gls{ice}) to zero-emission alternatives, such as battery electric (\gls{bev}), hydrogen-powered (\gls{fcev}) or zero-emission fuel-powered vehicles (e.g. running with biofuels or synthetic fuels). Adoption of light-duty electric vehicles is already gaining momentum globally, with 14 million sold in 2023 -- 60\% in China and 25\% in Europe, where the \gls{bev} stock reached 6.7~million with 2.2~million new registrations \cite{IEA_Global_EV_Outlook_2024}. Adoption rates vary significantly between light-duty and heavy-duty vehicles: in 2023, battery electric passenger cars accounted for 15\% of sales in the EU \cite{ACEA_Car_Registrations_2023}, while heavy trucks -- responsible for roughly 70\% of heavy-duty emissions \cite{EU_CO2_HDVs_2023} -- had just a 1\% share \cite{ICCT_Race_to_Zero_2024}. Despite these trends, \gls{bev}s made up only around 1.7\% of the total vehicle fleet in the European Union in 2023 \cite{eurostat2024ev}.

To accelerate the transition to zero-emission vehicles, CO$_2$ fleet emission standards for new vehicles in the EU and the UK require gradual emission reductions. Zero-emission light-duty vehicles are mandated in both the EU and the UK from 2035 onward, a target that was formerly envisioned for 2030 but delayed by five years \cite{GovUK_Zero_Emission_2023, EC_Fit_for_55_2022}. For heavy-duty vehicles, the EU mandates a 90\% reduction in emissions by 2040 compared to a reference year between 2019 and 2025, depending on the vehicle type \cite{EU_CO2_HDVs_2023}. Furthermore, the \gls{EU} has set the target of having 30~million zero-emission cars on the road by 2030 (roughly corresponding to 10\% of the current light-duty vehicle fleet) \cite{eu_mobility_strategy_2021}.

While numerous studies have explored decarbonisation pathways for road transport, most focus narrowly on only the transport sector, a single decarbonisation speed or a long-term scenario instead of the transition pathway. Electrification consistently emerges as the dominant strategy, with projected shares ranging between 40--100\% for road transport \cite{Ruhnau2019, Pickering2022, Bogdanov2024}. However, some studies allocate significant roles to hydrogen, biofuels, or synthetic fuels -- especially for heavy-duty vehicles -- with scenarios projecting up to 30\% \gls{fcev}s \cite{Luderer2022} or 27\% power-to-x fuels \cite{Helgeson2020} by 2050. The urgency of achieving net-zero emissions varies: while the EU Reference Scenario 2020 projects only a 35\% decrease of CO$_2$ emissions compared to 1990 \cite{Eu_reference_2020},  Plötz et al. \cite{Ploetz2023} argue that road transport emissions must reach net zero between 2044 and 2048 to stay within the \ce{1.5} carbon budget. The high-ambition scenarios for heavy-duty vehicles from an \gls{icct} report suggest even earlier emission reduction reaching net zero by 2040 \cite{icct_2023}.

Despite this growing body of research, an important gap remains: none of these studies examine how the speed of road transport decarbonisation affects other sectors. A report by the thinktank Agora \cite{agora_verkehrswende_2024} explores three decarbonisation pathways for Germany, ranging from fast transitions by 2040 to slower scenarios with residual emissions in 2050, finding that a faster transition is cost-optimal. However, this analysis is limited to a single country and focuses exclusively on the transport sector, leaving the broader implications for the overall energy systems unexplored.

This study addresses this gap by providing the first comprehensive analysis of how the pace of road transport decarbonisation in Europe shifts emission reduction burdens to other sectors assuming a fixed overall CO$_2$ budget. This is the first model with a high temporal and spatial resolution that endogenises road transport investment in light and heavy-duty as well as all other energy sectors. Using the sector-coupled energy system model PyPSA-Eur, we quantify the impact of faster or slower transport decarbonisation on renewable generation deployment, heat pump installations, green hydrogen production, and total system costs. Our model captures the competition for scarce resources (e.g., biomass), cross-sectoral interactions influencing fuel prices, competing decarbonisation of different sectors under a global budget and infrastructure requirements. Additionally, we assess the consequences of rising or declining transport demand and the benefits of maintaining zero-emission mandates for new light-duty vehicles by 2030, as initially planned in the EU and the UK, and heavy-duty vehicles by 2035.

\section{Methods}
\subsection{Model structure}
This study uses the European energy system model \gls{pypsa}-Eur \cite{githubpypsaeur} comprising the sectors power, heating, industry, agriculture, aviation, shipping and land transport. It minimises total system costs, while optimising generation, storage, transmission, distribution capacity and dispatch. Europe is represented by 39 regions, with a planning horizon spanning from 2025 to 2050, divided into five investment periods (2025, 2030, 2035, 2040, and 2050). The model has perfect foresight with 3-hourly time resolution inside each investment year, but myopic foresight between the investment periods. This means that investments cannot anticipate future CO$_2$ budgets or energy prices. The model is already described in detailed in multiple studies \cite{Synergies2018, victoria2022, neumann2023, zeyen2023} and in this study only newly-introduced features and modelling assumptions are discussed.

CO$_2$ emissions are constrained in line with EU greenhouse gas targets to a 55\% reduction in 2030, 90\% reduction in 2040 and net-zero in 2050 compared to 1990 levels. This corresponds to cumulative emissions of approximately 35~Gt$_{\text{CO}_2}$ from 2022 to 2050. For the European countries included in this study, with a total population of 611~million (compared to the global population of 8~billion, representing a 7.6\% share), this equates to global emissions of roughly 460~Gt$_{\text{CO}_2}$, which aligns with a +1.7$^\text{o}$C warming scenario (490~Gt$_{\text{CO}_2}$ with 67\% likelihood \cite{IPCC2021}) assuming no additional net-negative emissions after 2050 to balance emissions overshoot.

 CO$_2$ sequestration potentials are assumed to increase from zero in 2025 to 50~Mt$_{\text{CO}_2}$/a in 2030, reaching 200~Mt$_{\text{CO}_2}$/a by 2050. In addition, land use, land-use change, and forestry (\gls{lulucf}) emission reductions are available based on EU targets, ranging from 249~Mt$_{\text{CO}_2}$/a in 2025 to 310~Mt$_{\text{CO}_2}$/a from 2030 onward \cite{eu2023regulation_amending}. 77\% of this \gls{lulucf} potential is assumed available for mitigating CO$_2$ emissions (reflecting the share of CO$_2$ emissions relative to total greenhouse gas emissions in the EU-27 in 2021), while the remainder is allocated to offset other greenhouse gases. In practice, this means that \gls{lulucf} can reduce CO$_2$ emissions by up to 192~Mt$_{\text{CO}_2}$/a by 2025 and 239~Mt$_{\text{CO}_2}$/a from 2030 onward with costs of 25~\eurt \cite{ipcc2000lulucf}.
\begin{table*}[h!]
	\centering
		\begin{tabular}{@{}lcc@{}}
			\toprule
			\textbf{Scenario}    & \textbf{\begin{tabular}[c]{@{}c@{}}CO$_2$ Emission Limits\\ in New Vehicle Sales\end{tabular}}                              & \textbf{\begin{tabular}[c]{@{}c@{}}Road Transport\\ Demand\end{tabular}}                           \\ \midrule
			\textbf{Base}        & fulfilling EU CO$_2$ fleet targets                                                            & increasing according to historical growth                                                                                           \\ \hline
			\textbf{Fast}        & \begin{tabular}[c]{@{}c@{}}up to 100\% zero-emission vehicles \\ in new sales from 2030 onward\end{tabular}  & same as Base                                                                                           \\
			\textbf{Slow}        & \begin{tabular}[c]{@{}c@{}}maximum share of zero-emission vehicles from 16\%/10\% \\ in 2025 to 100\%/90\%  in 2050 for light/heavy-duty \\ \end{tabular} & same as Base                                                                                          \\ \hline
			\textbf{High-Demand} & same as Base                                                                                                     & \begin{tabular}[c]{@{}c@{}}increasing demand compared to Base\\ from 5\% in 2025 up to +30\% in 2050\end{tabular}   \\
			\textbf{Low-Demand}  & same as Base                                                                                                     & \begin{tabular}[c]{@{}c@{}}decreasing demand compared to Base \\ from -5\% in 2025 up to -30\% in 2050\end{tabular} \\ \bottomrule
			\textbf{Mandate}       & \begin{tabular}[c]{@{}c@{}}Allow only sales of zero-emissions \\ vehicles from 2030 (light) and 2035 (heavy)         \\    \end{tabular}  & same as Base                                                                                           \\ \hline
		\end{tabular}%
	\caption{Overview of the six different scenarios. Assumptions for every investment period are displayed in the Appendix in Figures \ref{fig:scenariooverview} and \ref{fig:max_share}.}
	\label{tab:scenarios}
\end{table*}

\subsection{Road Transport}
Road transport is represented with a distinction between light and heavy-duty vehicles, each characterized by different total demand based on vehicle-kilometre driven from the \gls{jrc}-\gls{idees} \cite{jrc2021} and demand time-series using weekly characteristic profiles from the German Federal Highway Research Institute (\gls{bast}) \cite{BASt}. The total demand is assumed to grow in the base case according to historical growth (light-duty annual average increase of 1.4\%/year and heavy-duty 1\%/year).

Endogenously, the road transport demand can be supplied by three different drivetrain technologies: (i) battery electric (\gls{bev}), (ii) hydrogen fuel cell (\gls{fcev}) or (iii) internal combustion engine (\gls{ice}) vehicles. The latter can operate either on fossil, synthetic or electrobiofuel. Electrobiofuels combine biofuel and electrofuel production processes, both based on Fischer-Tropsch synthesis, by adding renewable hydrogen to enhance carbon utilisation from biomass, resulting in higher fuel yields and improved efficiency \cite{Millinger2021}. Vehicle investment costs and efficiencies are taken for light-duty from \cite{fraunhofer_light_duty_cost} and for heavy-duty from the Danish Energy Agency \cite{DEA} assuming an average of capital costs across different truck types (see Figure \ref{fig:heavy-duty-investmentcosts} and Table \ref{tab:vehicle_tech_assumptions} in Appendix) due to the lack of detailed data on the share of different truck types in each country.

The fleet turnover is based on country-specific new car registration data from the \gls{jrc}-\gls{idees} database \cite{jrc2021} (see Appendix Figure \ref{fig:car_registration}), where the share of new registrations from the total fleet is endogenously optimised to determine the cost-optimal powertrain. Scenario-specific assumptions further define the maximum share of non-zero-emission vehicles from the newly registered cars. The total number of vehicles is assumed to stay constant over time. Since energy infrastructure needs are driven only by total vehicle-kilometre demand and not fleet size, while system costs are influenced by car investment costs, this assumption allows for a clearer distinction between the two factors. It is assumed that \gls{bev}s and \gls{fcev}s are used until the end of their lifetime, while \gls{ice}s can be retired earlier, but the annualised investment costs still have to be paid until the end of the vehicle's lifetime.

In the main scenarios, no demand side management (\gls{dsm}) in form of flexible charging of the \gls{bev} or vehicle-to-grid (\gls{v2g}) is allowed. As both are not yet widely implemented, excluding them serves as a cautious baseline to ensure results are not overly optimistic in favour of electrification. The option of \gls{dsm} and \gls{v2g} is explored as sensitivity in the Appendix and results in lower energy storage capacities for hydrogen and stationary batteries, lower capacities of the electricity distribution grid, and marginal lower system costs, cumulating to savings of  9 and 130~billion \officialeuro\ respectively over the whole modelling horizon, assuming a social discount rate of 2\% (see Section \ref{sec:sensi_v2g}).

This study does not explicitly model hybrid vehicles, gas-powered powertrains (\gls{cng}/\gls{lng}), and differentiation between petrol and diesel options. It also does not include further differentiation within heavy- and light-duty transport, such as long-haul trucks, buses, and two-wheelers. Infrastructure costs for charging are omitted, but a $\pm$20\% variation in vehicle capital costs is introduced to test the robustness of results.

\subsection{Scenarios}
We explore six scenarios that differ in the transition speed of road transport, the demand development, and if a mandate forbidding non-zero-emission vehicles from 2030 for light-duty and 2035 for heavy-duty is applied (see Table \ref{tab:scenarios}).

The \textbf{Base} scenario follows the EU-mandated fleet CO$_2$ emission reduction goals. For 2025, a modest increase in the share of zero-emission vehicles compared to 2023 levels is assumed, setting the maximum share for new vehicle sales at 20\% for light-duty and 15\% for heavy-duty vehicles. From 2030 onward, the maximum share of zero-emission vehicles follows the EU fleet reduction targets, such as 55\% for light-duty and 45\% for heavy-duty vehicles in 2030. From 2035 onward, up to 100\% zero-emission vehicles are allowed for light-duty (see Figure \ref{fig:max_share}). While these targets until 2035 could also be met by improving the efficiency of internal combustion engine vehicles, this study assumes no efficiency improvements for passenger cars, as historical trends show that vehicles' growing size and weight have offset efficiency increases. In the \textbf{Fast} transition scenario, more rapid emission reductions than politically targeted are reached, allowing up to 100\% zero-emission vehicles from 2030. In the \textbf{Slow} transition scenario, the CO$_2$ fleet targets are not met, and the adoption of zero-emission vehicles reaches up to 100\% of newly registered cars for light-duty only in 2050 and 90\% for heavy-duty.

To evaluate the impact of rising or decreasing demand compared to the historical trends, the \textbf{Low-demand} and \textbf{High-demand} scenarios assume changes in the road transport demand measured in driven kilometres relative to the \textbf{Base} scenario, starting at 5\% in 2025 and incrementally increasing by 5\% every five years, culminating in a total deviation of 30\% by 2050. For instance, in the \textbf{Low-demand} scenario, demand is 0.95 times the \textbf{Base} in 2025, 0.9 in 2030, and so forth (see Figure \ref{fig:scenariooverview}). In the \textbf{Mandate} scenario, no internal combustion engines can be registered from 2030 for light-duty and 2035 for heavy-duty onward.


\FloatBarrier
\section{Results}
\begin{figure*}[h!]
	\centering
	\begin{subfigure}[t]{0.45\textwidth}
		\includegraphics[width=\linewidth]{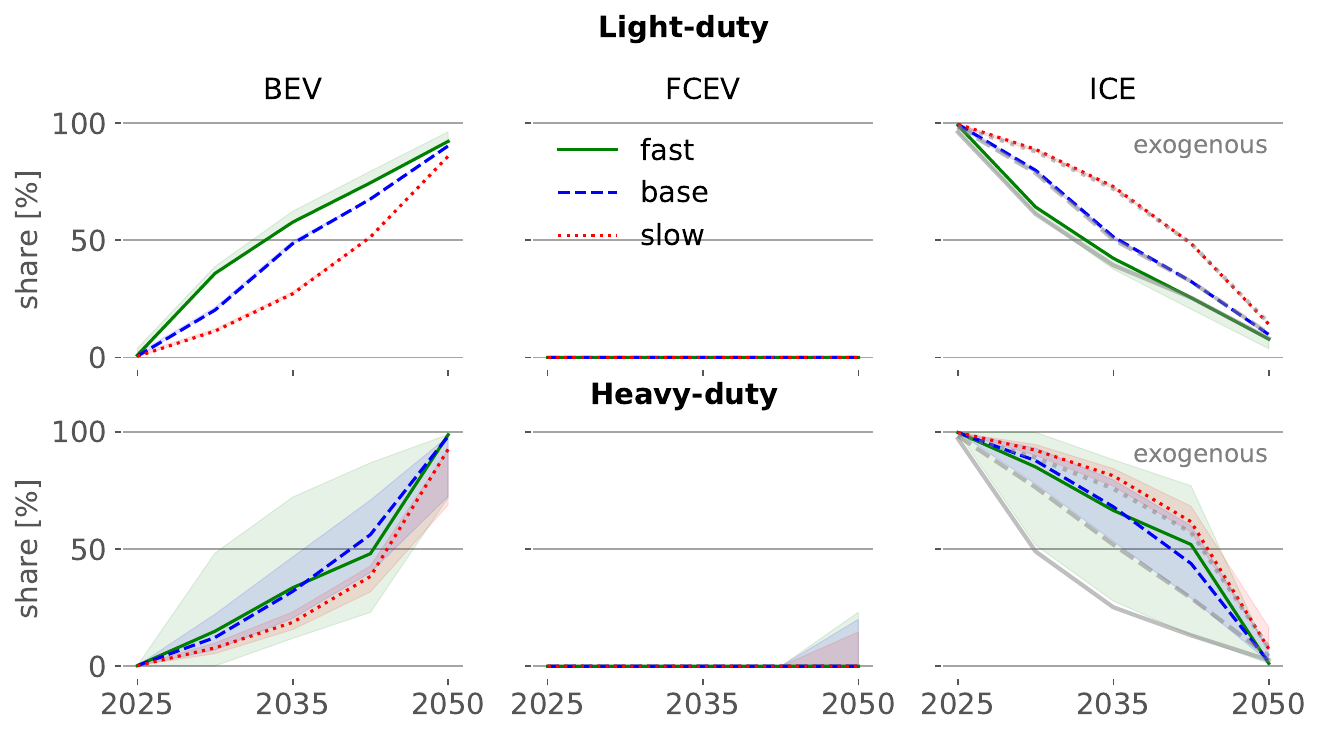}
		\caption{Variations of investment costs by $\pm$10\%}
		\label{fig:small_variations}
	\end{subfigure}
		\hfill
	\begin{subfigure}[t]{0.45\textwidth}
		\includegraphics[width=\linewidth]{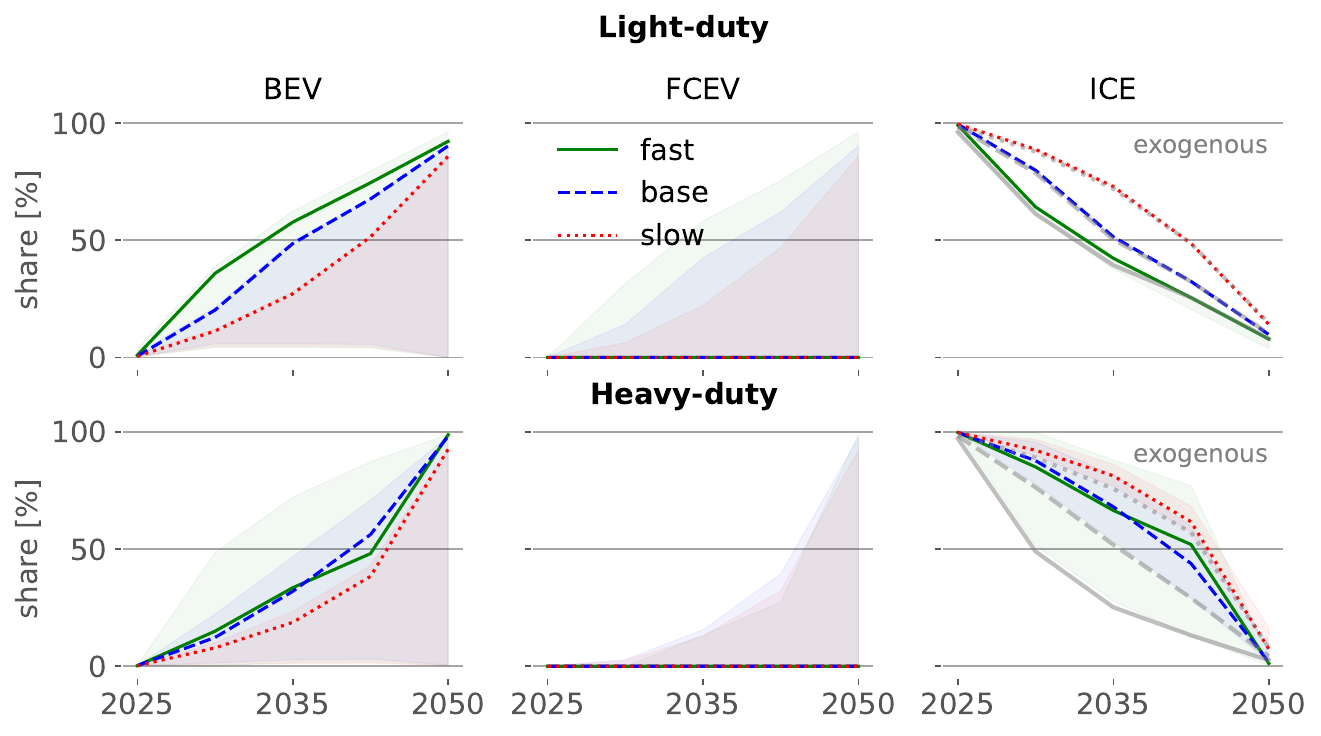}
		\caption{Variations of investment costs by $\pm$20\%}
		\label{fig:large_variations}
	\end{subfigure}	
	\caption{Share of engine type split by light and heavy-duty. Contour area indicates a variation of the vehicle investment costs by $\pm$10\% and $\pm$20\%.  These variations reflect scenarios where, for instance, the investment cost of \gls{bev}s is increased by 10\% while the cost of \gls{fcev}s is simultaneously decreased by 10\%, and vice versa.}
	\label{fig:area-share}
\end{figure*}

In all scenarios, the electrification of both light and heavy-duty vehicles is the cost-optimal strategy to decarbonise road transport. \gls{fcev}s are only adopted under significant investment cost reductions for \gls{fcev}s combined with increased investment costs for \gls{bev}s (see Figure \ref{fig:area-share}). The option of producing electrobiofuels is never used and synthetic fuel production starts in 2040, but mainly to decarbonise aviation and shipping.

The EU target of 30~million zero-emission vehicles (roughly corresponding to a minimum share of 10\% zero-emission vehicles in light-duty) in 2030 is reached in all scenarios. A faster transition in the light-duty vehicle sector enables a slower transition for heavy-duty vehicles where earlier decarbonisation is associated with higher costs due to higher zero-emission truck costs in the early years.

All scenarios demonstrate the need for an accelerated scale-up of renewable energy capacities (see Figure \ref{fig:annual_additional_res} in Appendix) and, in most scenarios, a faster deployment of heat pumps (see Figure \ref{fig:annual_additional_heat_pumps} in Appendix) compared to historical values.

\subsection{Rapid Electrification of Light-Duty Reduces Pressure on Fast Scale-up of Renewable, Heat Pump and Electrolysis Installation}
For light-duty vehicles, the switch from \gls{ice} to zero-emission vehicles is constraint by the assumed fleet turnover rate. For the \textbf{Slow} and \textbf{Base} scenario, this constraint is binding, also under significant variations of vehicle investment costs. In contrast, a slower transition for heavy-duty than light-duty is cost-optimal, and the allowed fleet turnover is not constraining. The option of retiring \gls{ice} before the end of their lifetime is not used since the investment costs for the vehicles are so high. Once the investment decision for one powertrain is made, the vehicle is used until the end of its lifetime. To evaluate the impact of this limited foresight, we investigate the advantages of having earlier political mandates for allowing only zero-emission vehicles in new sales in Section \ref{sec:mandates}.

\begin{figure}[h!]
	\centering
	\includegraphics[width=\linewidth]{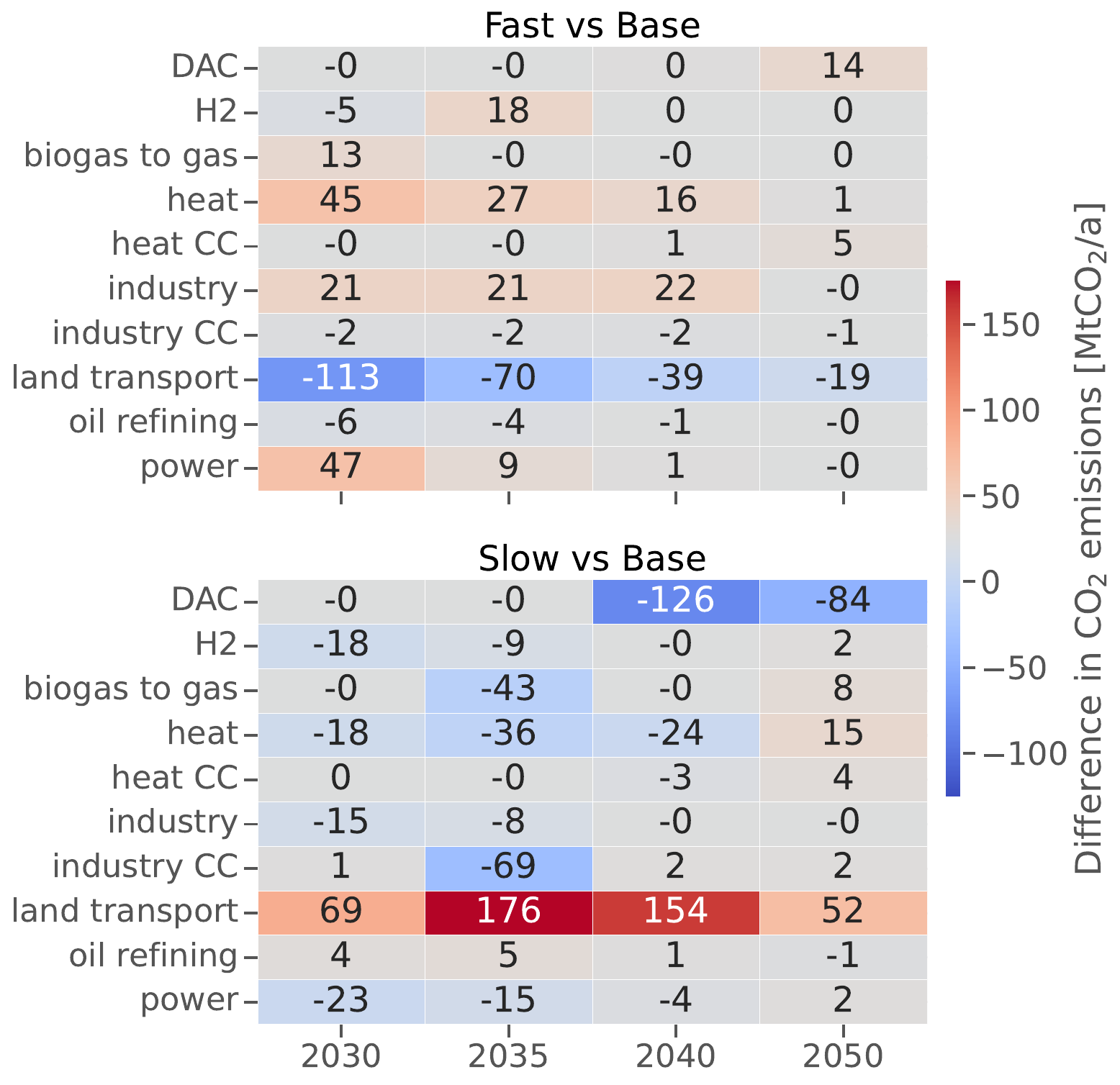}
	\caption{Difference in CO$_2$ emissions compared to the \textbf{Base} scenario.}
	\label{fig:heatmap_co2_emissions}
\end{figure}

The pace of decarbonisation of road transport impacts the transition speed in the other sectors and the total system costs (see Figure \ref{fig:heatmap_co2_emissions}). A faster transition of road transport allows for a slower build-out of renewable generation capacities, installation of heat pumps, ramp up of green hydrogen production, and reduces reliance on carbon capture. 

\begin{figure*}[h!]
	\centering
	\includegraphics[width=\linewidth]{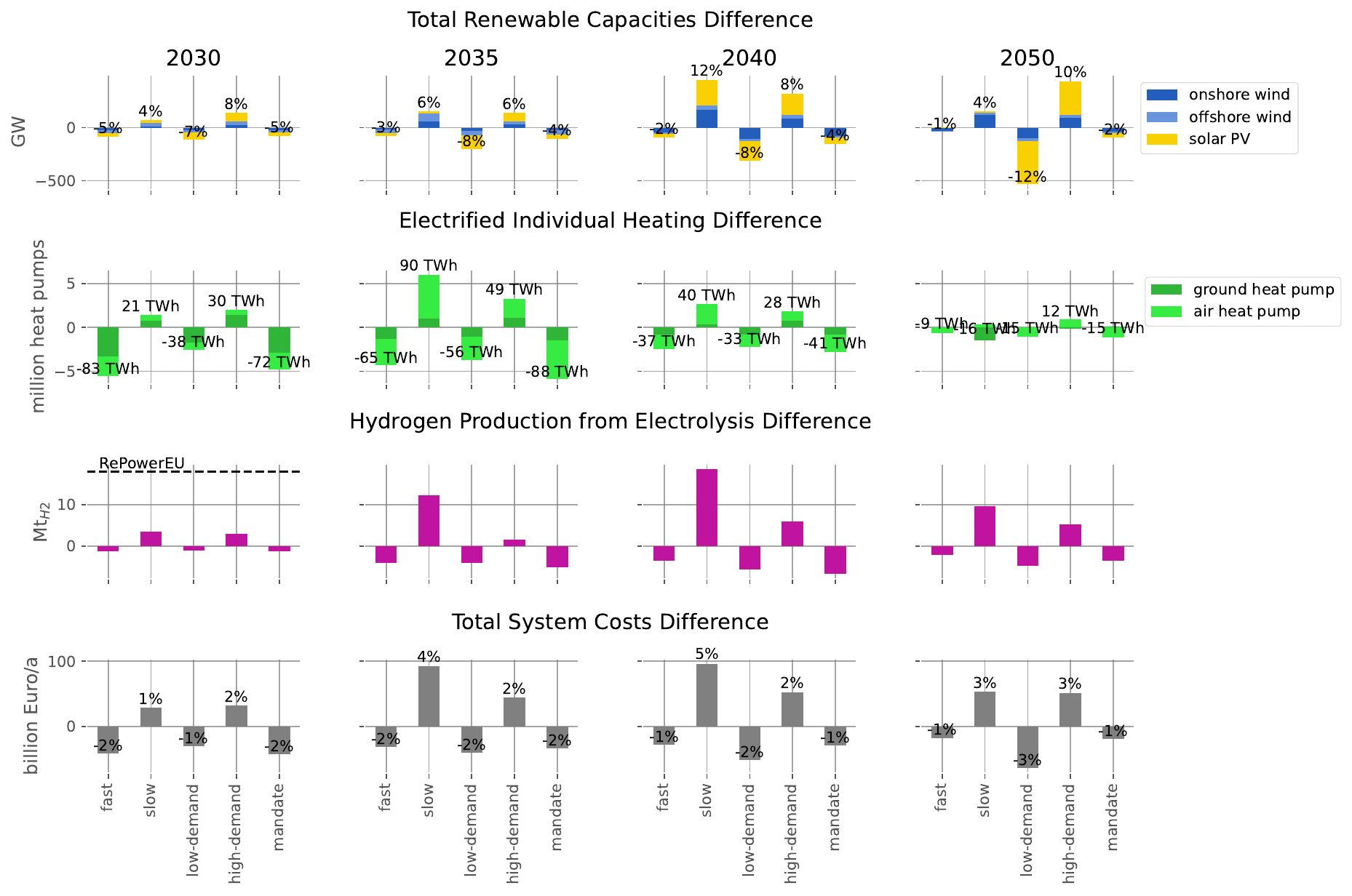}
	\caption{Key factors compared to the \textbf{Base} scenario.}
	\label{fig:comparison}
\end{figure*}

In the near term (2025--2035), power, heating and industry compensate for varying levels of emissions in road transport to meet climate targets. A slower transition in road transport requires until 2035 an up to 22\% higher additional annual renewable build-out rate (162--169~GW/a) compared to a fast transition (138--147~GW/a) (see Figure \ref{fig:comparison}). This is a significant challenge, since all scenarios necessitate at least doubling the historical annual build-out rate of 70~GW in 2022, with total renewable capacities reaching 2.3--2.6~TW by 2035 (Figures \ref{fig:installed_res} and \ref{fig:annual_additional_res}). A fast transition reduces the need for renewable capacities since it is more efficient to electrify the road transport first compared to other sectors, e.g. production of green hydrogen can be delayed.

In the \textbf{Slow} scenario, the delayed electrification of road transport shifts a significant share of the emission reduction burden to the heating sector. This necessitates the installation of 3.6~million additional heat pumps annually between 2025--2030, assuming a typical generation of 15~MWh$_\text{th}$/a per heat pump. This demand represents a 9\% increase compared to the \textbf{Base} scenario and a 64\% rise above historical installation rates. In contrast, the \textbf{Fast} scenario enables a more gradual electrification of the heating sector. Annual installations drop to 2.2~million heat pumps, which is even below the historical level of 2.9~million heat pumps in 2021 \cite{EHPA2023}. 

The scale-up of green hydrogen production is in all scenarios below the political target of producing 10~Mt$_{\text{H}_2}$ in 2030 within Europe envisioned by RePowerEU \cite{repowereu}. Since with a slower transition in road transport there are fewer remaining free emissions, in the near-term a faster transition from grey to green hydrogen production is required with 55~GW installed electrolysis capacities in 2030 (5~Mt$_{\text{H}_2}$  produced) in the \textbf{Slow} scenario versus 7~GW in the \textbf{Fast} scenario (with 0.7~Mt$_{\text{H}_2}$). In the long-term, more synthetic fuel needs to be produced in the \textbf{Slow} scenario due to the higher share of \gls{ice}, which need to run with zero-emission fuels and therefore require larger capacities of electrolysis and renewable generation. Since this is less efficient than electrifying road transport, total system costs are up to 5\% (98~billion \officialeuro/a) higher compared to the \textbf{Base} scenario. On the contrary, with a \textbf{Fast} transition, especially in the near-term, up to 2\% (43~billion \officialeuro/a) compared to the \textbf{Base} scenario can be saved. 

In 2030 system-wide CO$_2$ prices are more than two times higher in the \textbf{Slow} transition scenario (290\eurt) compared to the \textbf{Fast} scenario (137\eurt) and hydrogen prices about 41\% higher (110\mwhh versus 78\mwhh). This is caused by more costly mitigation of CO$_2$  in other sectors and the need to ramp up the green hydrogen production earlier, when higher exogenous costs are assumed.

In the long-term (2040--2050), higher CO$_2$ emissions from road transport are mitigated by direct air capture (\gls{dac}) in the \textbf{Slow} scenario, capturing up to 126~Mt$_{\text{CO}_2}$/a in 2040. This higher reliance on carbon capture with a slow transition in road transport depends on uncertain factors such as the (i) future costs of \gls{dac}, (ii) availability of a CO$_2$ network and (iii) CO$_2$ sequestration potentials \cite{hofmann2024}. In the long-term, \textbf{Fast} and \textbf{Base} scenarios result in similar shares of electric vehicles. The slightly higher emissions in the \textbf{Base} scenario are offset by faster electrification in heating and decarbonisation of industry.

\subsection{Higher Road Transport Demand Drives Up Total System Costs and Renewable Deployment}
In the previous scenarios, vehicle-kilometres driven are assumed to increase annually based on historical growth trends. This section presents a sensitivity analysis exploring the impact of gradually increasing or decreasing road transport demand, ranging from  $\pm$5 in 2025 to $\pm$30\% in 2050 relative to the \textbf{Base} scenario. The analysis isolates the impact of vehicle-kilometres driven and assumes a constant vehicle fleet size, which accounts for a significant fraction of roughly 60\% of total system costs. Independent of road transport demand levels, electrification is the preferred strategy for decarbonising both heavy- and light-duty vehicles, which remains the most efficient and cost-effective approach.

Higher road transport demand increases total system costs, primarily due to the increased need for renewable energy generation capacity and the accelerated decarbonisation of other sectors, which is often more costly. By 2050, the \textbf{High-demand} scenario requires an additional 313~GW of renewable capacity compared to the \textbf{Base} scenario, driving up system costs by 3\% (53~billion \officialeuro/a). In contrast, reduced transport demand alleviates these pressures, with 408~GW less renewable capacity needed and total system costs 65~billion \officialeuro/a lower than in the \textbf{Base} scenario. These findings highlight a key distinction from the previous scenarios, which explored varying transition speeds for decarbonising road transport. This analysis underscores that the level of road transport demand has a long-term impact even after complete decarbonisation. Higher transport demand drives greater electricity consumption, necessitating substantially more renewable generation capacity to sustain the system. Conversely, lower demand reduces the renewable capacity requirements, easing system costs and potential public acceptance problems.

In addition to lowering costs, reduced transport demand and connected emissions slow the pace of decarbonisation required in other sectors. Between 2025 and 2030, annual renewable capacity additions in the \textbf{Lower-demand} scenario are 14\% lower (134~GW/a) than in the \textbf{Base} case, easing the strain on a fast deployment. Lower demand leads to a slower ramp-up of green hydrogen production, decreasing from 11 to 7~Mt$_{\text{H}_2}$ versus 13~Mt$_{\text{H}_2}$ with higher demand in 2030, and a reduced deployment of heat pumps, with 2.9~million additional heat pumps between 2025--2030 (-13\% compared to \textbf{Base}) versus 3.6~million annual additional heat pumps (+8\% compared to \textbf{Base}) with increasing demand. Since in 2040--2050 there are some remaining emissions from the road transport, in the \textbf{High-demand} scenario \gls{dac} is used to capture up to 37~Mt$_{\text{CO}_2}$/a more compared to the \textbf{Base} scenario, while with a lower-demand the reliance is reduced by 21~Mt$_{\text{CO}_2}$/a (see Figure \ref{fig:co2-heatmap-demand}).

\subsection{Avoiding Costly Lock-In: Benefits of Earlier Zero-Emission Vehicle Mandates}\label{sec:mandates}
Early retirement of \gls{ice} is avoided because the high vehicle investment costs incentivize usage over the entire lifetime. In the model, the decisions on the powertrain are based in the model on energy prices at the time of investment without the foresight of future years and changes in fuel prices. Evaluating costs over the entire operational lifetime of vehicles could lead to another vehicle type being more cost-optimal in the long run. We perform a sensitivity analysis to address this potential mismatch and prevent investment in more costly assets over the whole lifetime. This analysis evaluates the potential benefits of implementing political mandates requiring all newly-registered light-duty vehicles to be zero-emission by 2030, as initially planned by the EU, and heavy-duty vehicles by 2035, thus preventing stranded assets.

Since light-duty BEVs become cost-competitive from 2030 onward,  the 2035 zero-emission mandate is primarily relevant for accelerating the transformation in the heavy-duty sector. Further, a mandate would not only ensure road transport transformation but also eliminate uncertainties by guaranteeing the adoption of zero-emission vehicles, even if vehicle costs are not reducing as fast as expected, and offer manufactures planning security. 

Compared to current policies, the earlier \textbf{Mandate} scenario for zero-emission vehicles saves annual costs of up to 2\% (43~billion \officialeuro/a), which cumulates over the whole modeling horizon to 533~billion \officialeuro, assuming a 2\% social discount rate. Also compared to the \textbf{Fast} transition scenario, which allows for zero-emission vehicles of up to 100\% in new vehicle sales from 2030 onward, stranded investments in internal combustion vehicles in heavy-duty in 2035 are prevented, which results in cumulative cost-savings of 14~billion \officialeuro (assuming a social discount rate of 2\%). The mandates further allow for a slower deployment of renewable generation capacity, e.g. between 2035--2040 the annual build-out rates of additional renewable capacity are 265~GW, 3--4\% lower compared to the \textbf{Fast} and \textbf{Base} scenario. Reduced emissions from heavy-duty transport especially allow for a slower decarbonisation of the heating sector with 5.9 and 1.7~million and  heat pumps fewer installed in 2035 compared to \textbf{Base} and \textbf{Fast} scenario, respectively.
\FloatBarrier
\section{Discussion and Conclusion}
This study examines how the pace of road transport decarbonisation impacts the burden on other sectors to achieve emission reductions to stay within the European climate targets. Our findings emphasize that electrification -- for both light and heavy-duty vehicles -- is the dominant strategy for decarbonisation, while alternative fuels such as biofuels and hydrogen remain largely untapped even under large vehicle investment cost variations. This aligns with findings from other literature \cite{icct_2023, agora_verkehrswende_2024, Bogdanov2024, Ruhnau2019}.

We demonstrate that a fast transition of road transport not only minimises overall costs and results in a lower CO$_2$ price but also alleviates the urgency for rapid deployment of renewable energy capacity, heat pumps, and electrolysis, as well as reducing reliance on carbon capture technologies. This conclusion is consistent with Agora's study \cite{agora_verkehrswende_2024}, which focuses on Germany and similarly finds that a faster transition is more cost-effective, even when analysed from a transport-sector-specific perspective. However, even with a fast transition in road transport, our results show that the annual deployment rate of renewable energy capacities must at least double compared to the historical rate, and green hydrogen production needs to be scaled up to 7~Mt$_{\text{H}_2}$ by 2035 to meet the political emission reduction targets. Breed et al. \cite{breed2021}, focusing on heavy-duty transport in Europe with more detailed fleet modelling than in our study, project 4--22\% zero-emission truck sales and 2--11\% of total stock in 2030 under previous CO$_2$ emission targets (-30\% by 2030 compared to 2019/2020). Under the updated targets (-45\% by 2030 compared to 2021--2024), we find slightly higher values of 8--15\% zero-emission vehicles in the total heavy-duty stock by 2030. Similarly, the BloombergNEF Electric Vehicle Outlook 2024 \cite{bnef_ev_outlook_2024} estimates a 41\% share of electric vehicles in new passenger car sales in Europe by 2027, which aligns between our Base and Slow scenarios.

A key driver of emissions and cost reductions is the trajectory of transport demand. In our model, the Low-demand scenario assumes a reduction in the vehicle-kilometres traveled while the number of vehicles is kept constant. This could represent a future scenario in which passenger-kilometres are reduced for instance thanks to home-office, use of public transport or bicycles some days of the week. It could also represent a scenario in which the traveled passenger-kilometres increase but the number of passengers per vehicle also increases thanks to car-sharing or targeted policies.  Reducing the vehicle-kilometres travelled has significant additional benefits which are not included in our modelling,  such as lower particle pollution from fine particulates, reduced number of accidents, and safer and healthier cities. Our results indicate that curbing the historical growth by up to 30\% by 2050 could yield annual savings of up to 65~billion \officialeuro/a. Reducing road transport demand reduces the need for additional renewable generation capacity and reliance on \gls{dac} in the long term. This further avoids land-use and public acceptance problems and mitigates uncertainties surrounding the development of carbon capture technologies. Similarly, the analysis by Kany et al.  \cite{Kany2022} analysis of the Danish transport sector decarbonisation shows that reducing transport demand can lower transport sector costs by 10\% while achieving full decarbonisation by 2045. Furthermore, Deblas et al. \cite{Deblas2020} show that reductions in road transport demand are essential for staying within a \ce{1.5--2} global warming threshold. We show that, in contrast, a slow transition in road transport places a high burden on other sectors for emission reductions to stay within the climate target, e.g., it demands the installation of 3.6~million additional heat pumps annually in the near-term, far exceeding the 2.9 million installed in 2021. 

We show that earlier mandates than envisaged by European policy -- requiring zero-emission light-duty vehicles in new sales from 2030 and heavy-duty vehicles from 2035 -- avoid investments in stranded assets and are cost-optimal compared to current policies. Our findings align with Plötz et al. \cite{Ploetz2023}, who investigated only the transport sector and suggest similar mandates from 2033 for light-duty and 2033--2038 for heavy-duty vehicles to meet stringent \ce{1.5} targets. In agreement with these results, BloombergNEF EV outlook \cite{bnef_ev_outlook_2024} advocates for policy interventions to ensure heavy-duty vehicle electrification, which remains critical even though they state that heavy-duty electric vehicles become economically viable from 2030 onward.

Taken together, this study demonstrates that fast electrification of road transport is critical for alleviating decarbonisation pressures on other sectors, reducing reliance on carbon capture in the long term and cost-effectively achieving European climate goals. To realise this potential, earlier EU mandates for zero-emission vehicles are essential to avoid stranded assets and ensure long-term economic efficiency. This strategy should be complemented by policies encouraging road transport demand reduction to allow for a more gradual transition in the other sectors.

\section*{Credit Author Statement}
\textbf{Elisabeth Zeyen}: Methodology, Software, Validation,
Formal analysis, Investigation, Data Curation, Writing - Original Draft, Visualization 

\textbf{Sina Kalweit}: Writing- Review \& Editing, Software 

\textbf{Marta Victoria}: Writing- Review \& Editing, Conceptualization 

\textbf{Tom Brown}: Conceptualization, Methodology, Resources, Software,
Writing- Review \& Editing, Supervision,  Project administration
\section*{Conflict of Interest Statement}
The authors declare no competing interests.

\section*{Data Availability Statement}
The modelling workflow and data behind this study are open. The entire project is available in a public repository under MIT license \cite{zenodo_report}. The repository also contains a summary of output data for all scenarios. The code is available in a Github repository \cite{githubtransport}. The technology data assumptions are from the open Github repository \texttt{technology-data} version 0.9.2 \cite{technologydata}.

\section*{Funding Statement}
 S.K. and M.V are partially funded by the Novo Nordisk CO2 Research Center (CORC) under grant number CORC005.
\clearpage
\FloatBarrier
\newpage

\printglossary[type=\acronymtype]
\FloatBarrier
\clearpage
\bibliography{transport}
\clearpage
\onecolumn
\section*{Appendix}
\begin{figure}[h]
	\centering
	\includegraphics[width=0.7\linewidth]{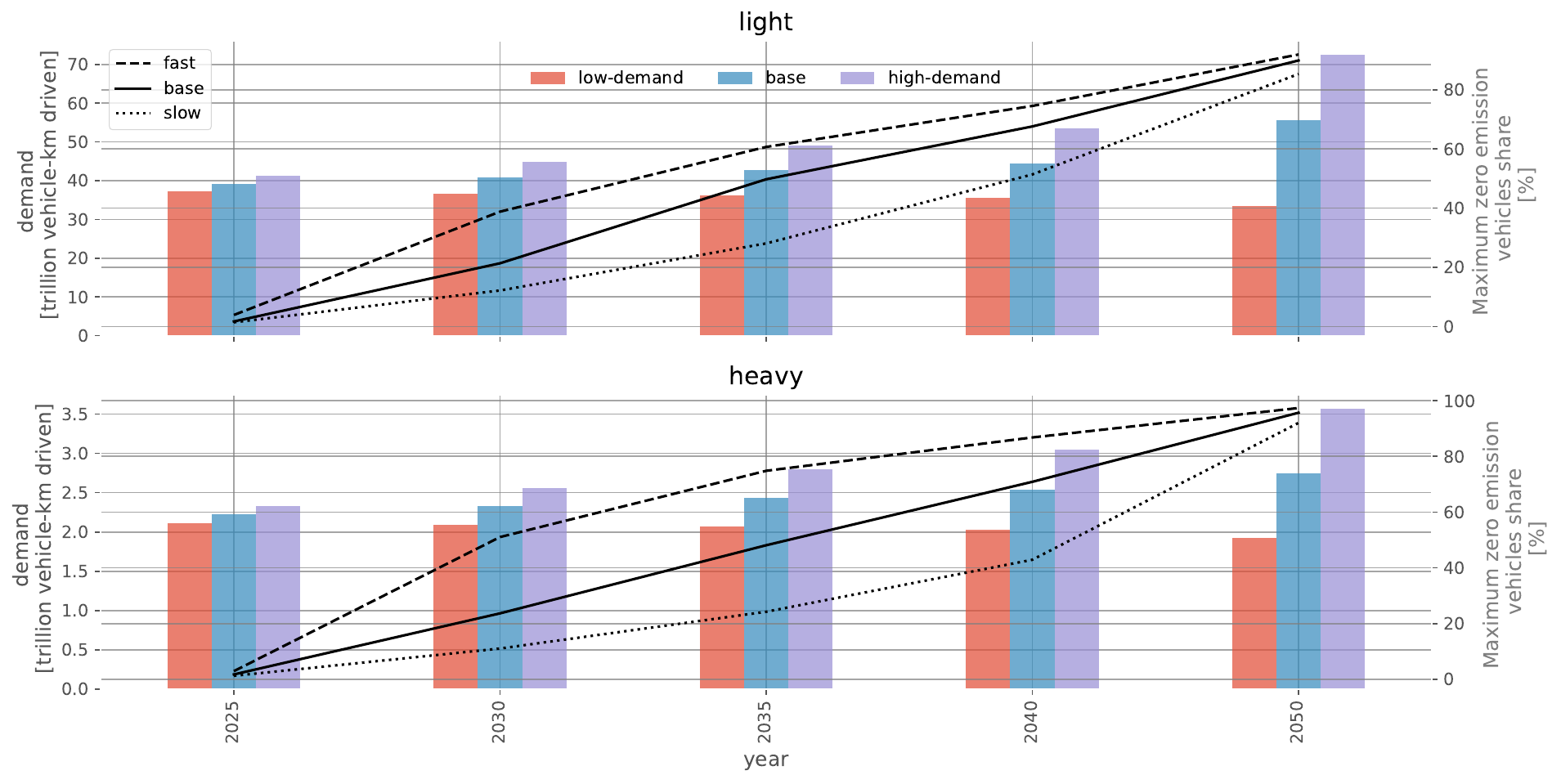}
	\caption{Overview of the scenarios.}
	\label{fig:scenariooverview}
\end{figure}

\begin{figure}[h]
	\centering
	\includegraphics[width=0.7\linewidth]{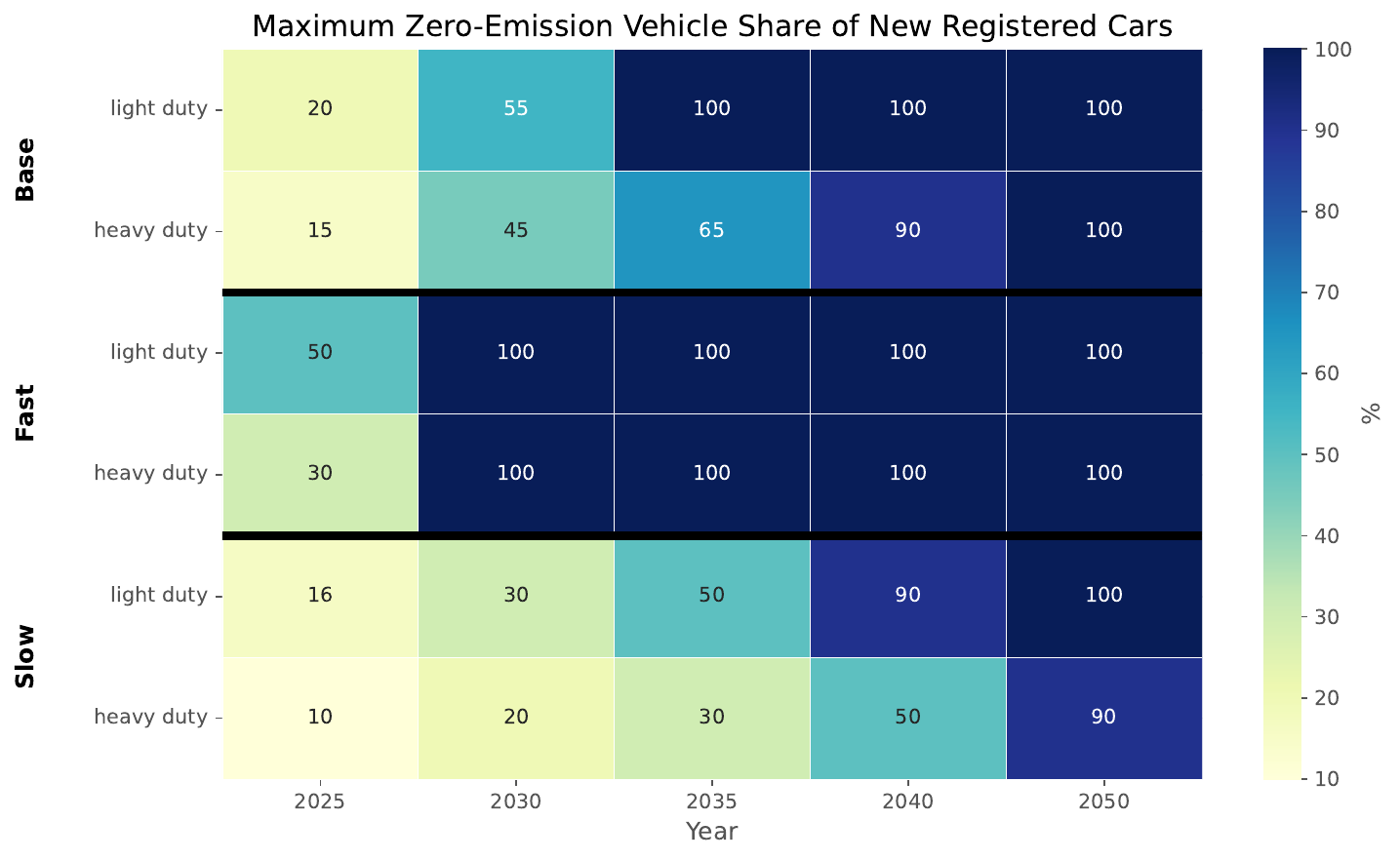}
	\caption{Maximum share of zero-emission cars from newly registered vehicles depending on the scenario.}
	\label{fig:max_share}
\end{figure}

\begin{figure*}[h!]
	\centering	
	\begin{subfigure}[b]{0.45\linewidth}
		\centering
		\includegraphics[width=\linewidth]{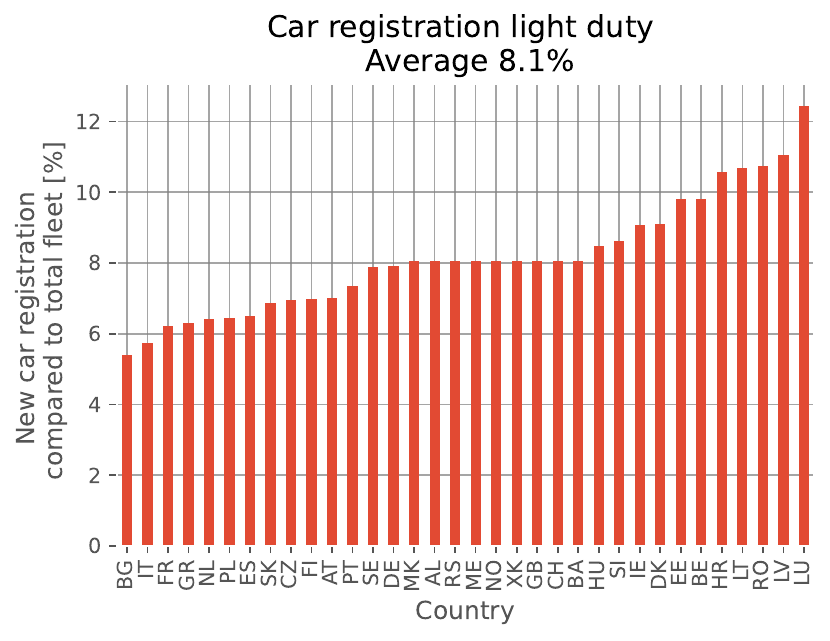}
		\caption{light-duty}
		\label{fig:car_registration_light_duty}
	\end{subfigure}
	\hfill
	\begin{subfigure}[b]{0.45\linewidth}
		\centering
		\includegraphics[width=\linewidth]{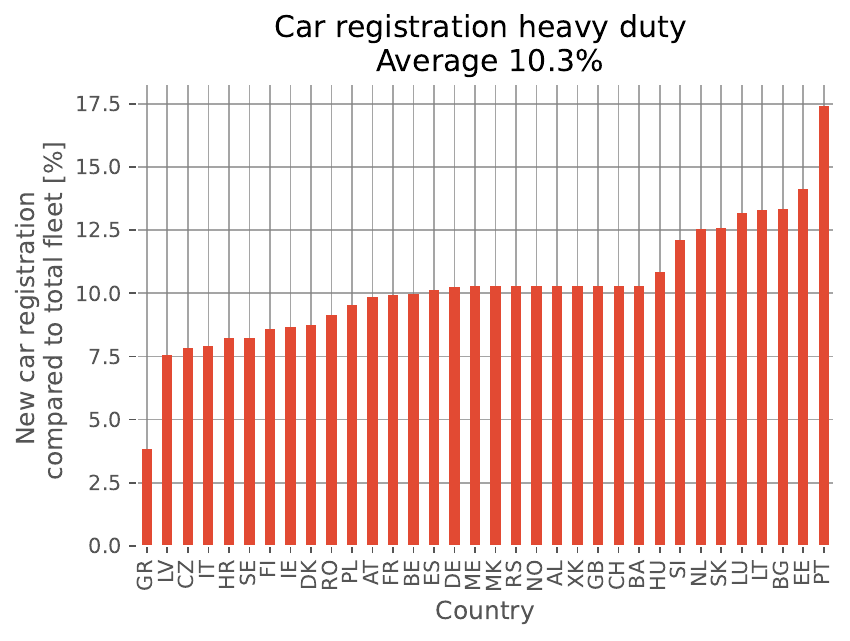}
		\caption{heavy-duty}
		\label{fig:car_registration_heavy_duty}
	\end{subfigure}	
	\caption{Historical demand for light and heavy-duty road transport based on \gls{jrc}-\gls{idees}-2021 \cite{jrc2021}.}
	\label{fig:car_registration}
\end{figure*}

\begin{figure*}[h!]
	\centering	
	\begin{subfigure}[b]{0.45\linewidth}
		\centering
		\includegraphics[width=\linewidth]{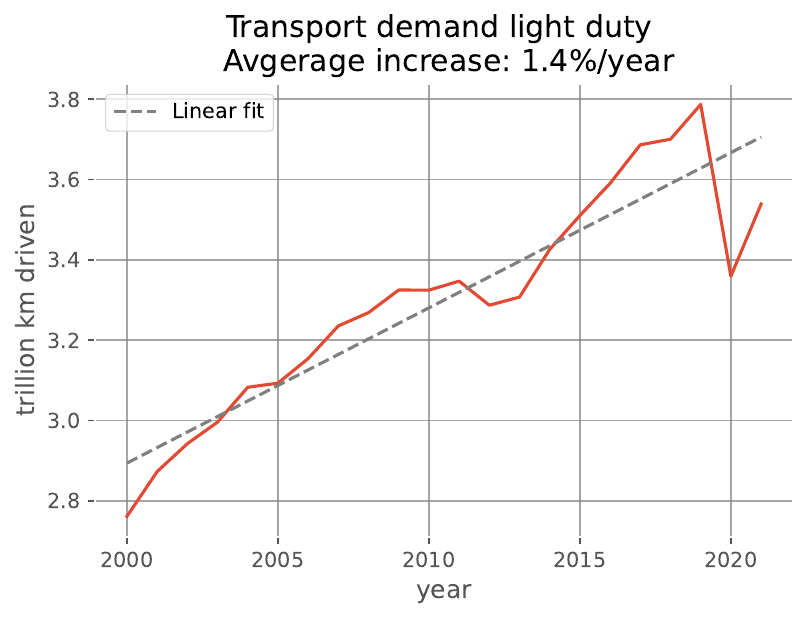}
		\caption{light-duty}
		\label{fig:demand_light_duty}
	\end{subfigure}
	\hfill
	\begin{subfigure}[b]{0.45\linewidth}
		\centering
		\includegraphics[width=\linewidth]{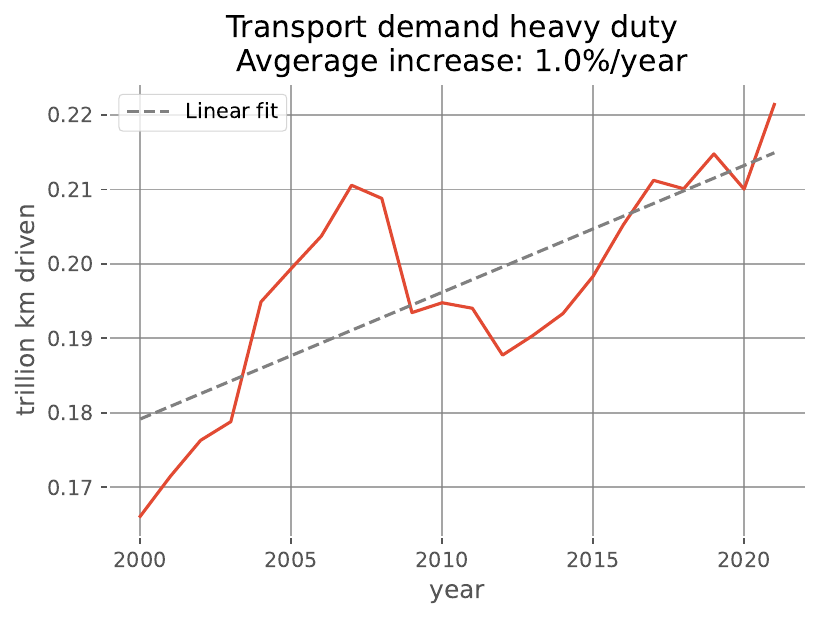}
		\caption{heavy-duty}
		\label{fig:demand_heavy_duty}
	\end{subfigure}	
	\caption{Historical demand for light and heavy-duty road transport based on \gls{jrc}-\gls{idees}-2021 \cite{jrc2021}.}
	\label{fig:demand_historical}
\end{figure*}

\begin{figure*}[h!]
	\centering	
	\begin{subfigure}[b]{0.45\linewidth}
		\centering
		\includegraphics[width=\linewidth]{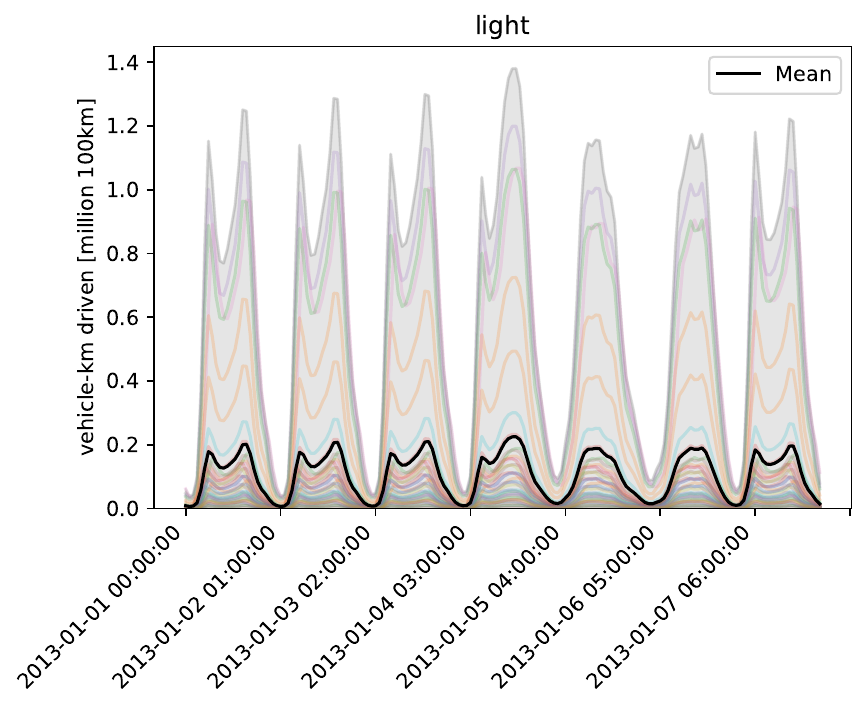}
		\caption{light-duty}
		\label{fig:transport_demand_example_week_light_duty}
	\end{subfigure}
	\hfill
	\begin{subfigure}[b]{0.45\linewidth}
		\centering
		\includegraphics[width=\linewidth]{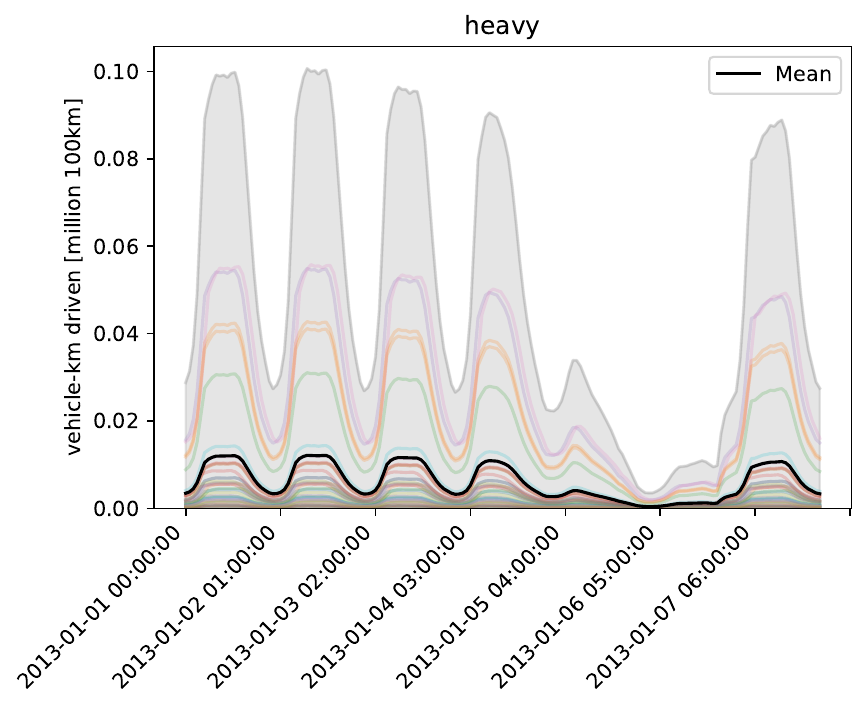}
		\caption{heavy-duty}
		\label{fig:transport_demand_example_week_heavy_duty}
	\end{subfigure}	
	\caption{Example week for transport demand time series.}
	\label{fig:transport_demand_example_week}
\end{figure*}

\begin{figure}
	\centering
	\includegraphics[width=0.5\linewidth]{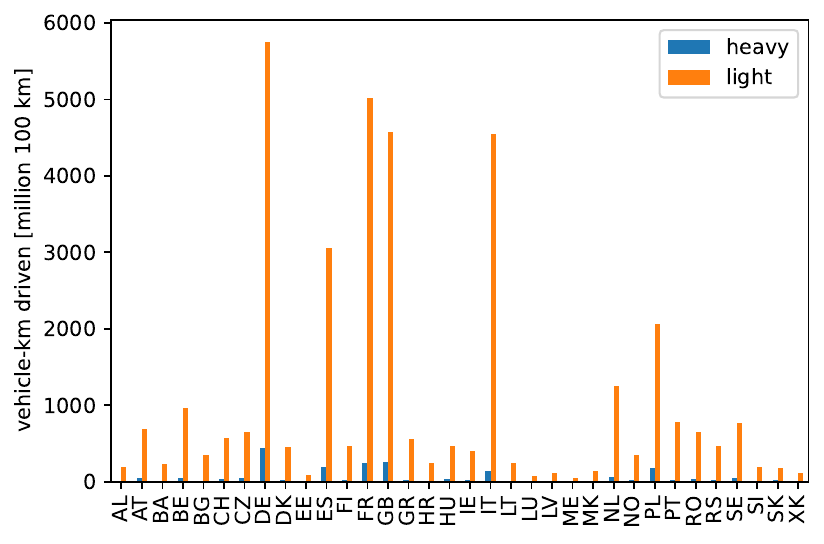}
	\caption{Summed vehicle-km driven for heavy and light-duty split by country}
	\label{fig:transport_demand_total}
\end{figure}

\begin{figure}
	\centering
	\includegraphics[width=0.7\linewidth]{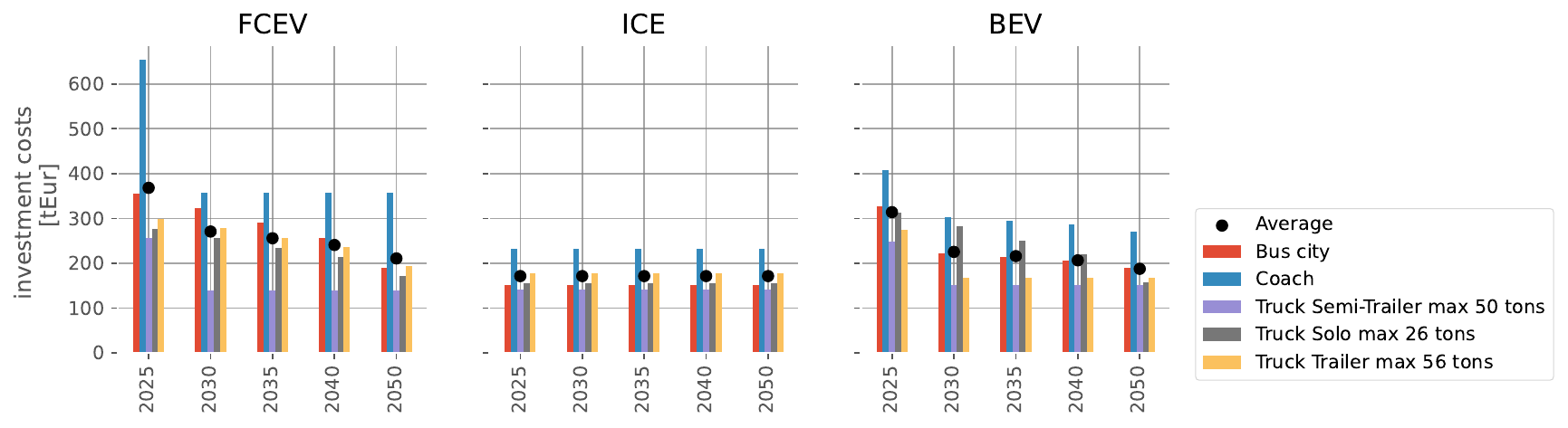}
	\caption{Investment costs for different vehicle and engine types and assumed average investment costs.}
	\label{fig:heavy-duty-investmentcosts}
\end{figure}

\begin{figure}[h!]
	\centering
	\includegraphics[width=\linewidth]{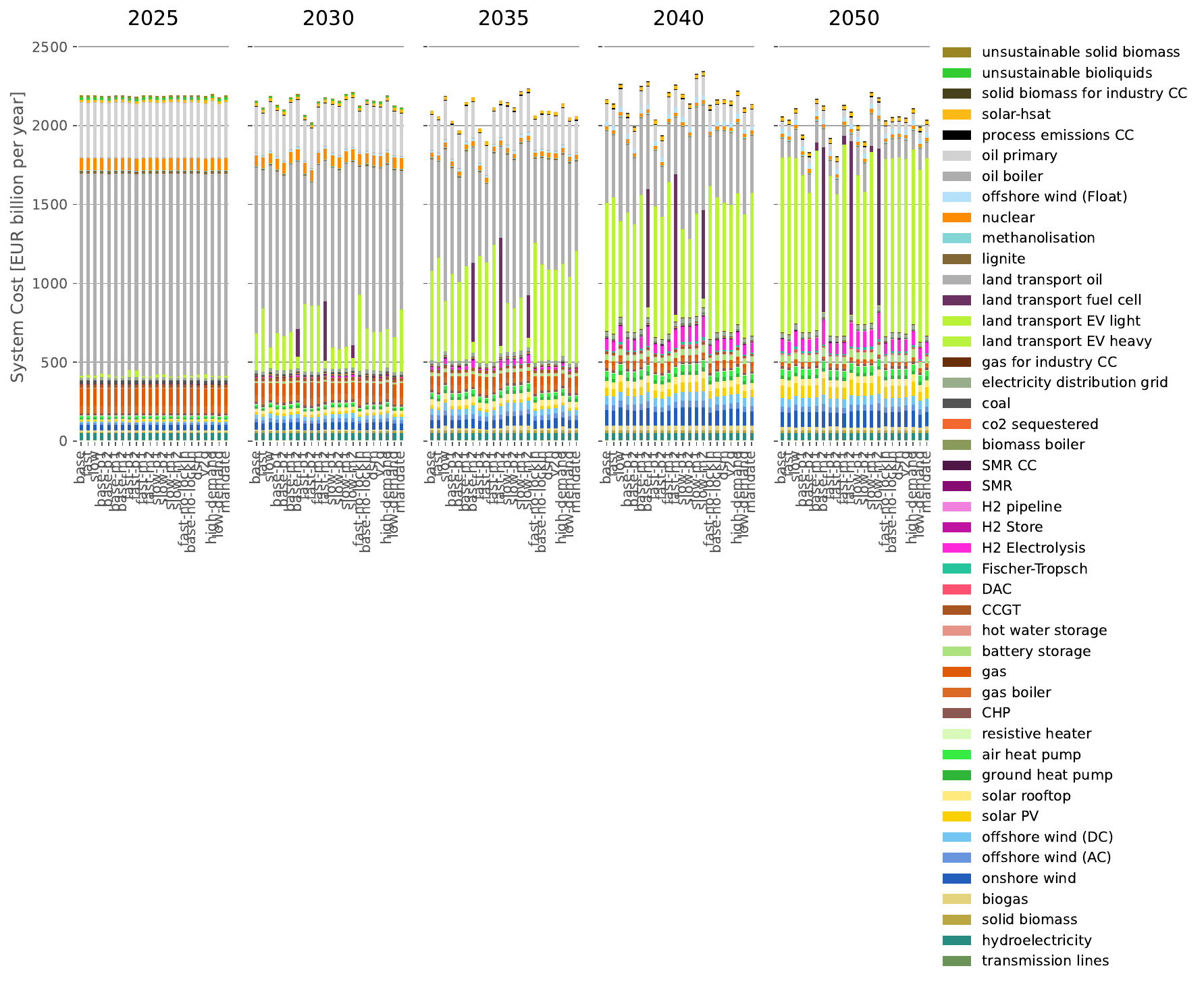}
	\caption{Total annualised system costs.}
	\label{fig:total_costs}
\end{figure}

\begin{figure*}[h!]
	\centering
	\begin{subfigure}[b]{0.45\linewidth}
		\centering
		\includegraphics[width=\linewidth]{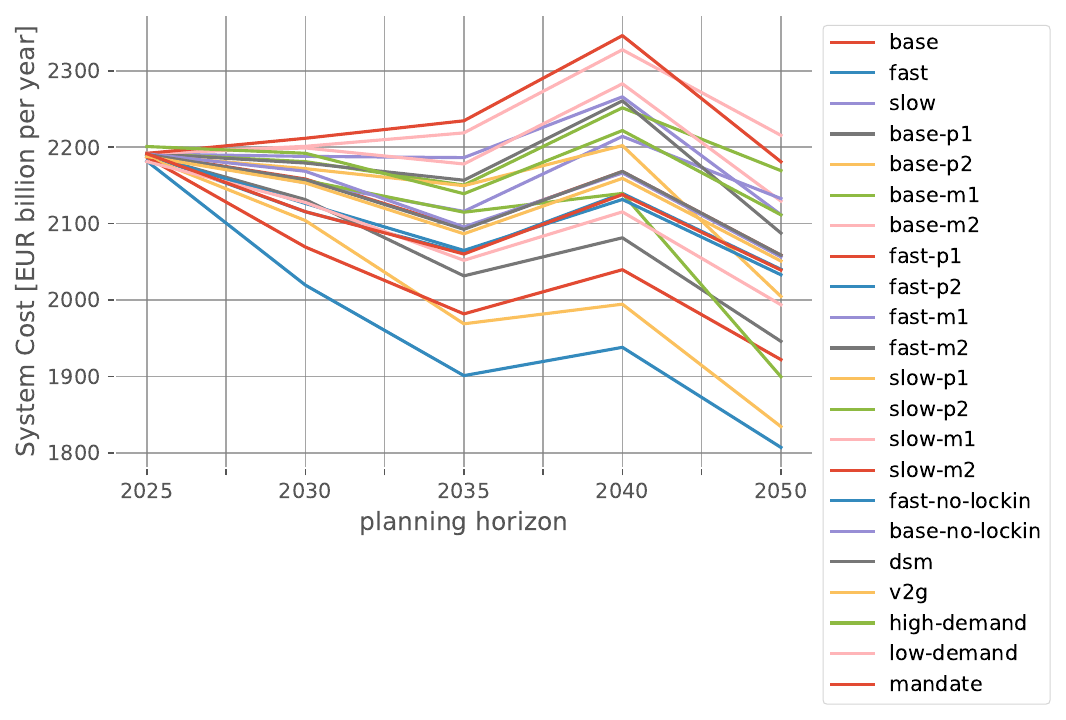}
		\caption{Summed total annualised costs.}
		\label{fig:total_costs_summed}
	\end{subfigure}
	\hfill
	\begin{subfigure}[b]{0.45\linewidth}
		\centering
		\includegraphics[width=\linewidth]{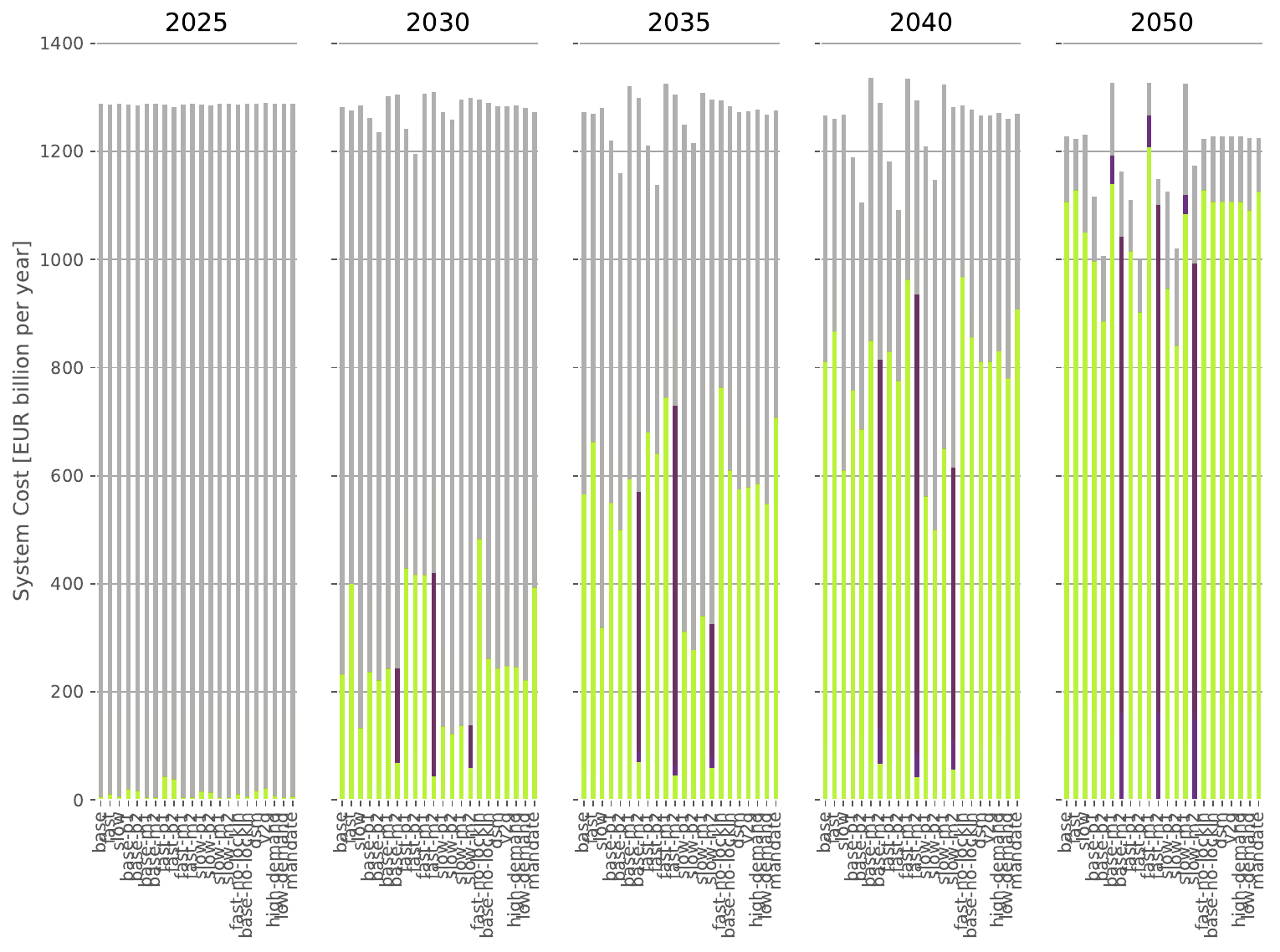}
		\caption{Annualised costs road transport.}
		\label{fig:cost_transport}
	\end{subfigure}
	
	\caption{Total summed annualised costs and costs of the road transport sector.}
	\label{fig:summed_costs}
\end{figure*}

\begin{figure*}[h!]
	\centering
	\begin{subfigure}[b]{0.45\linewidth}
		\centering
		\includegraphics[width=\linewidth]{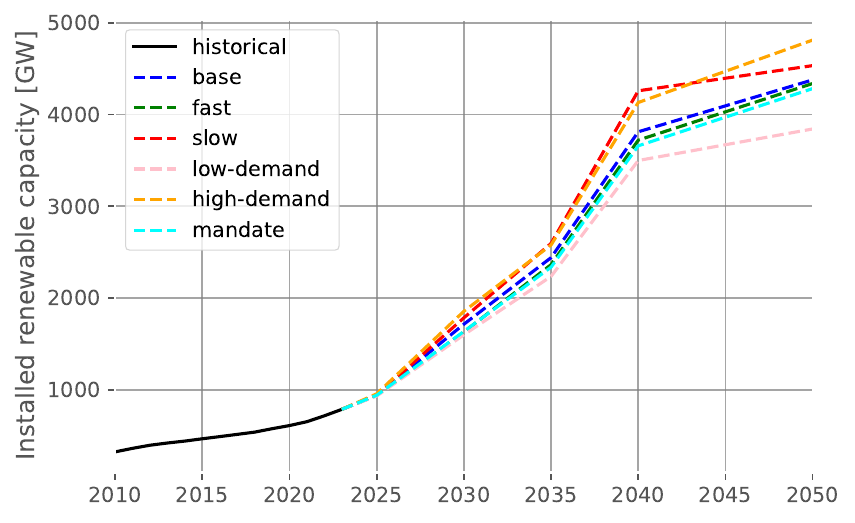}
		\caption{Installed renewable capacities.}
		\label{fig:installed_res}
	\end{subfigure}
	\hfill
	\begin{subfigure}[b]{0.45\linewidth}
		\centering
		\includegraphics[width=\linewidth]{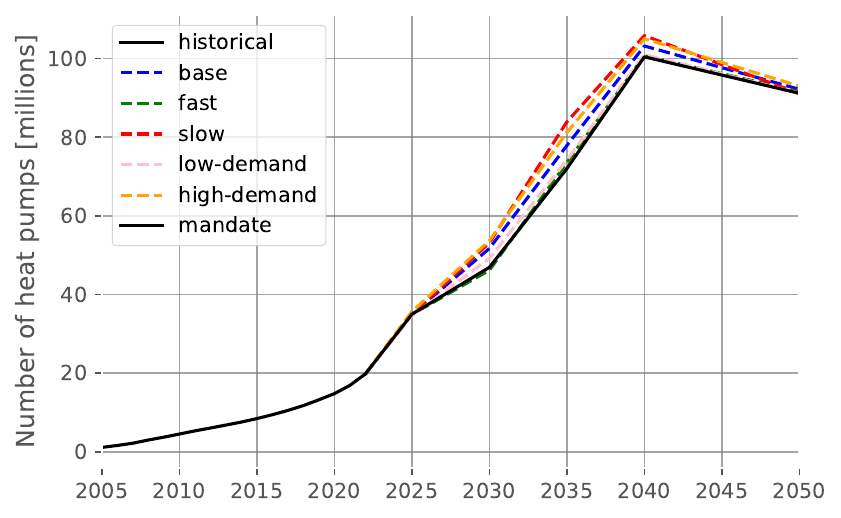}
		\caption{Number of installed heat pumps for individual heating.}
		\label{fig:installed_heat_pumps}
	\end{subfigure}
	
	\caption{Installed renewable capacities and number of installed heat pumps for individual heating.}
	\label{fig:installation_scenarios}
\end{figure*}

\begin{figure}
	\centering
	\includegraphics[width=0.7\linewidth]{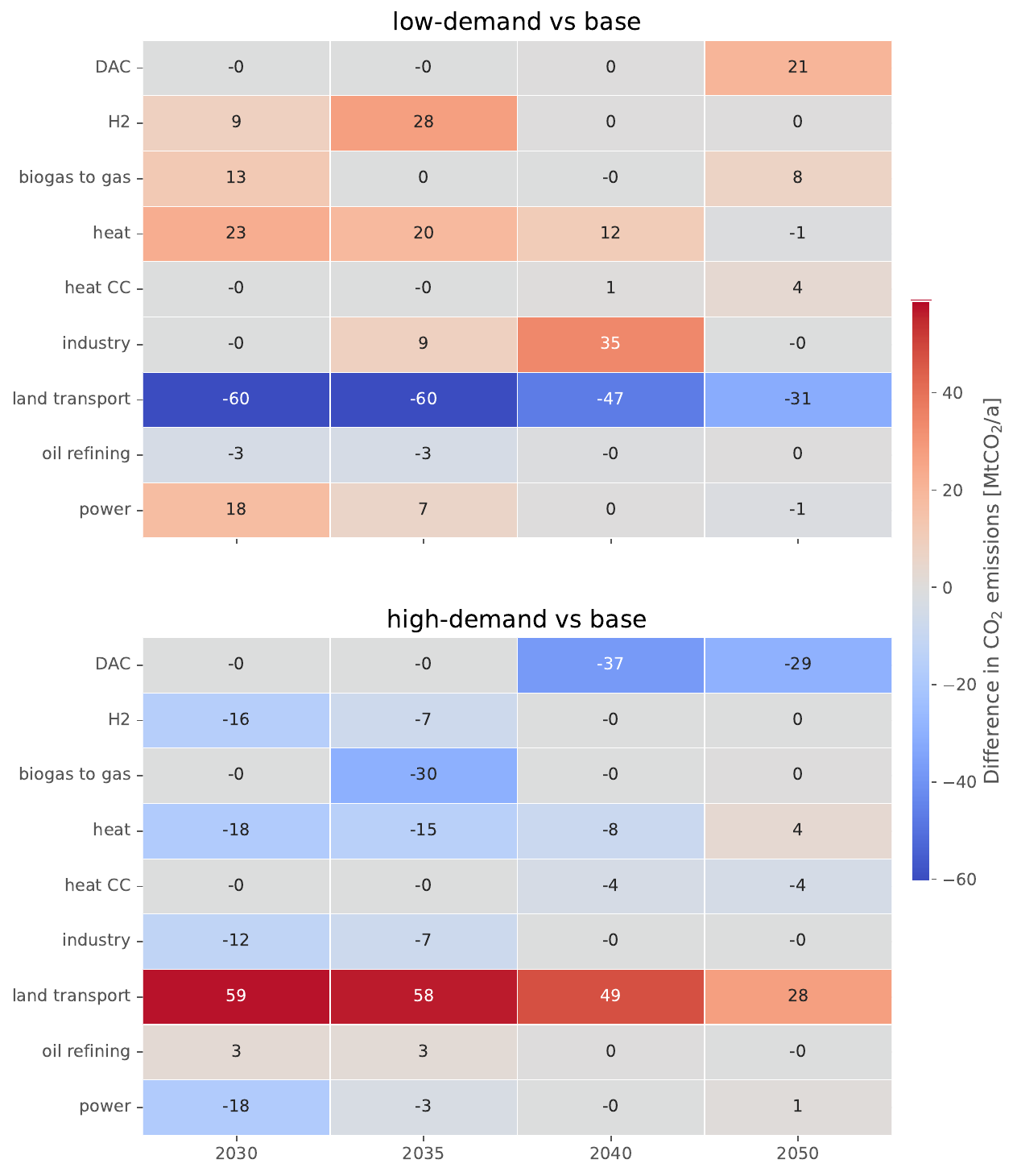}
	\caption{Difference in CO$_2$ emissions from \textbf{High-demand} and \textbf{Low-demand} scenario compared to \textbf{Base}.}
	\label{fig:co2-heatmap-demand}
\end{figure}

\begin{figure*}[h!]
	\centering
	\begin{subfigure}[b]{0.45\linewidth}
		\centering
		\includegraphics[width=\linewidth]{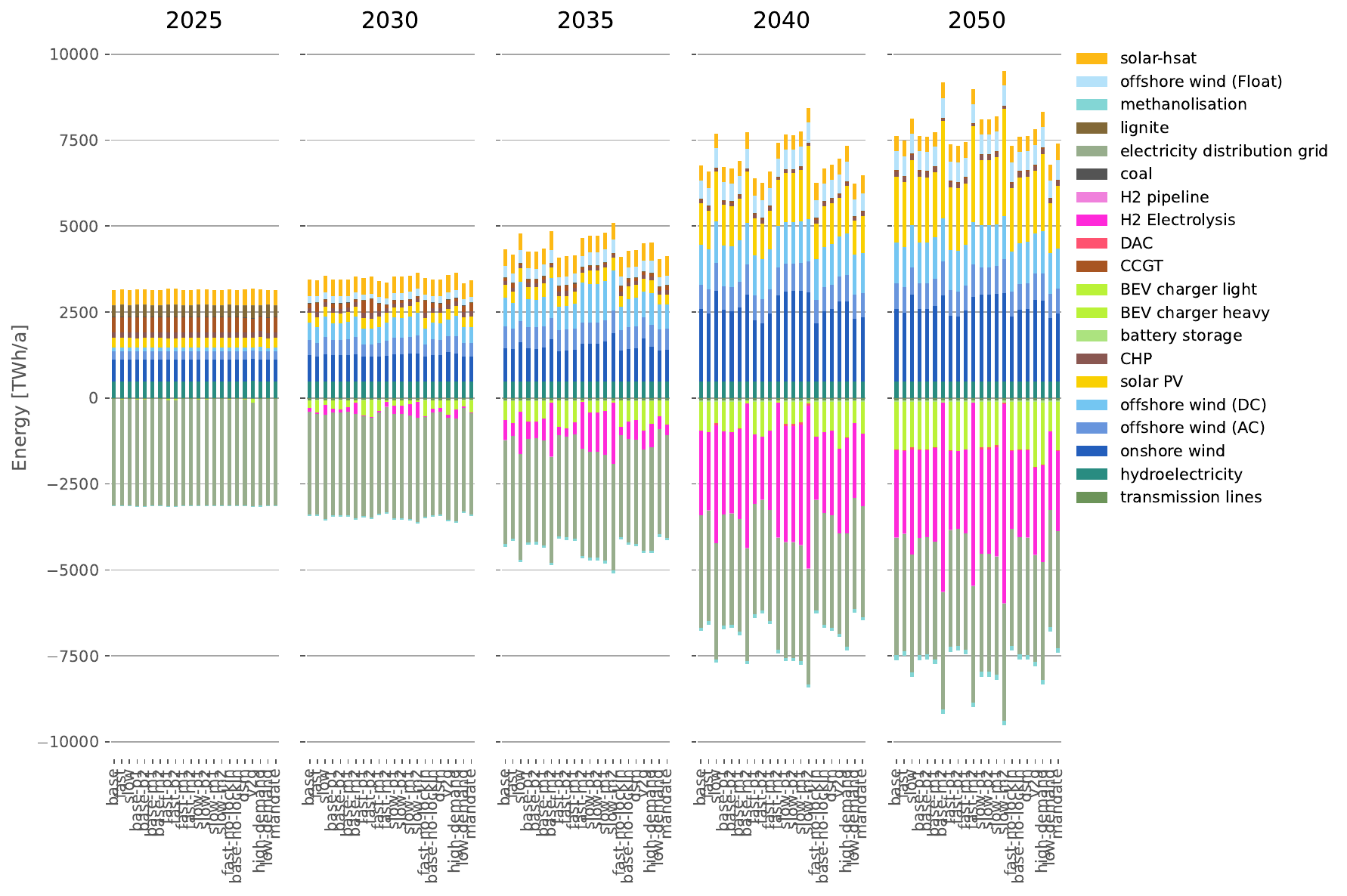}
		\caption{AC}
		\label{fig:balances_AC}
	\end{subfigure}
	\hfill
	\begin{subfigure}[b]{0.45\linewidth}
		\centering
		\includegraphics[width=\linewidth]{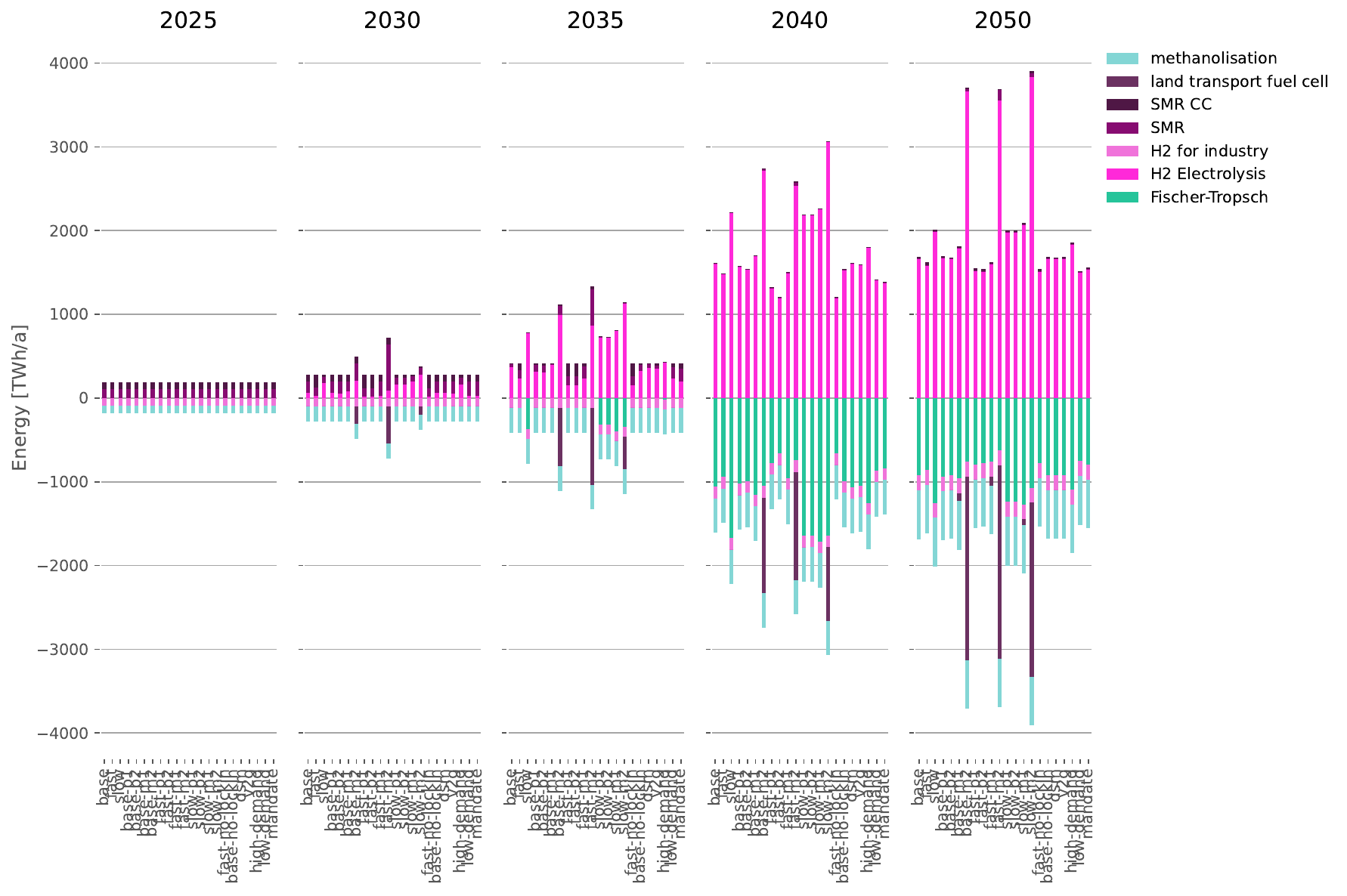}
		\caption{H$_2$}
		\label{fig:balances_H2}
	\end{subfigure}
	\begin{subfigure}[b]{0.45\linewidth}
		\centering
		\includegraphics[width=\linewidth]{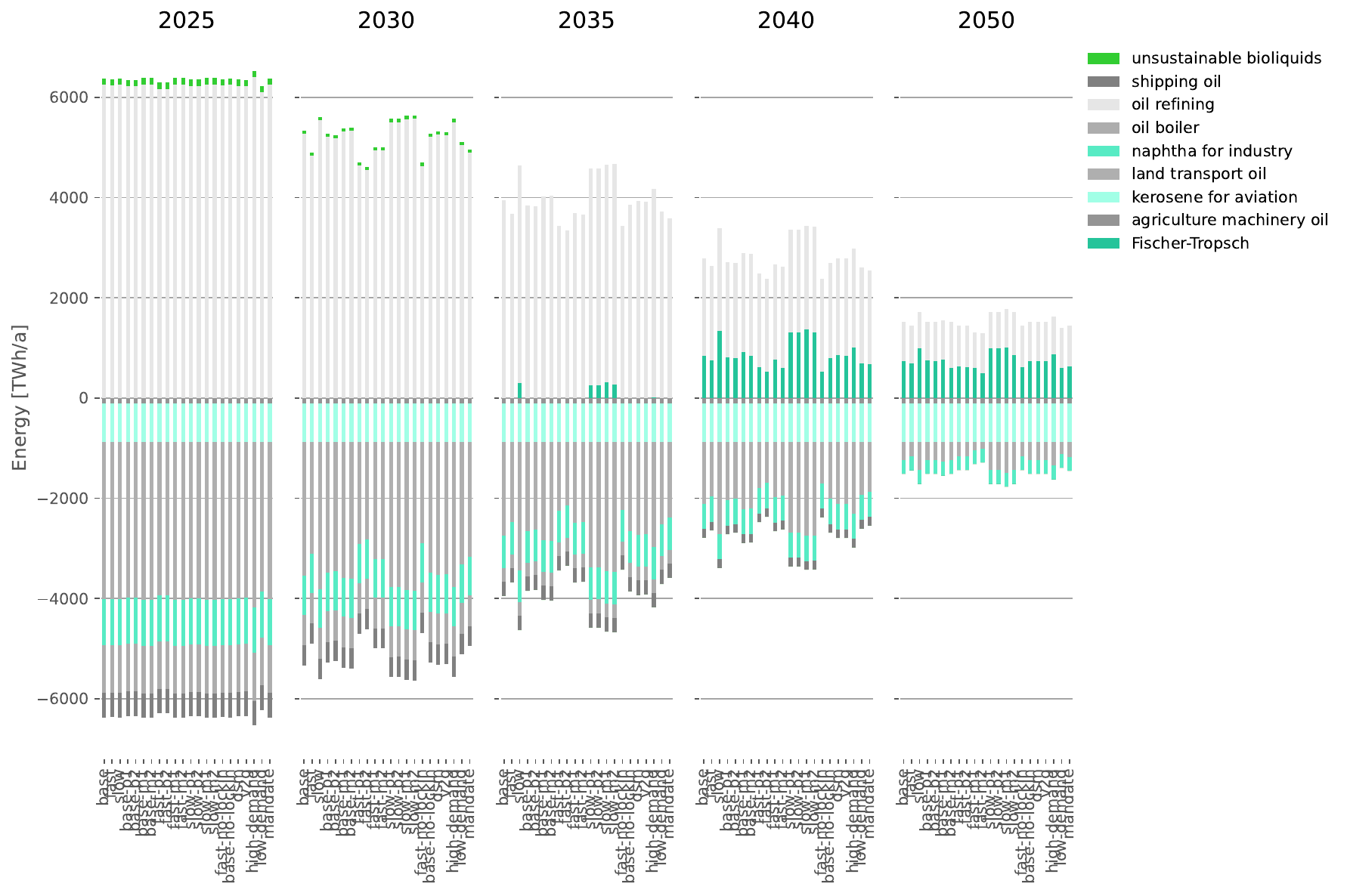}
		\caption{oil}
		\label{fig:balances_oil}
	\end{subfigure}
	\hfill
	\begin{subfigure}[b]{0.45\linewidth}
		\centering
		\includegraphics[width=\linewidth]{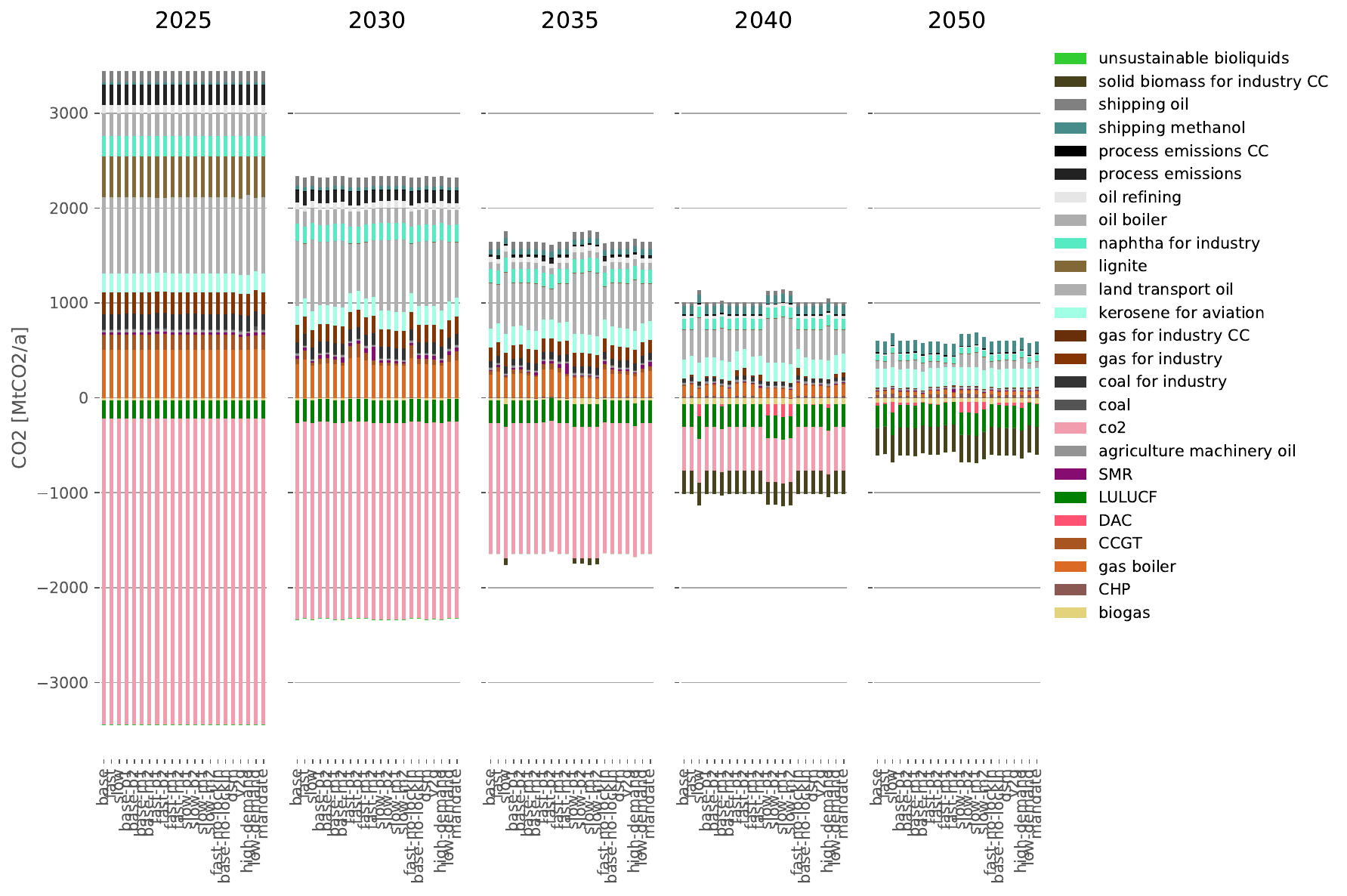}
		\caption{CO$_2$}
		\label{fig:balances_CO2}
	\end{subfigure}
	\begin{subfigure}[b]{0.45\linewidth}
		\centering
		\includegraphics[width=\linewidth]{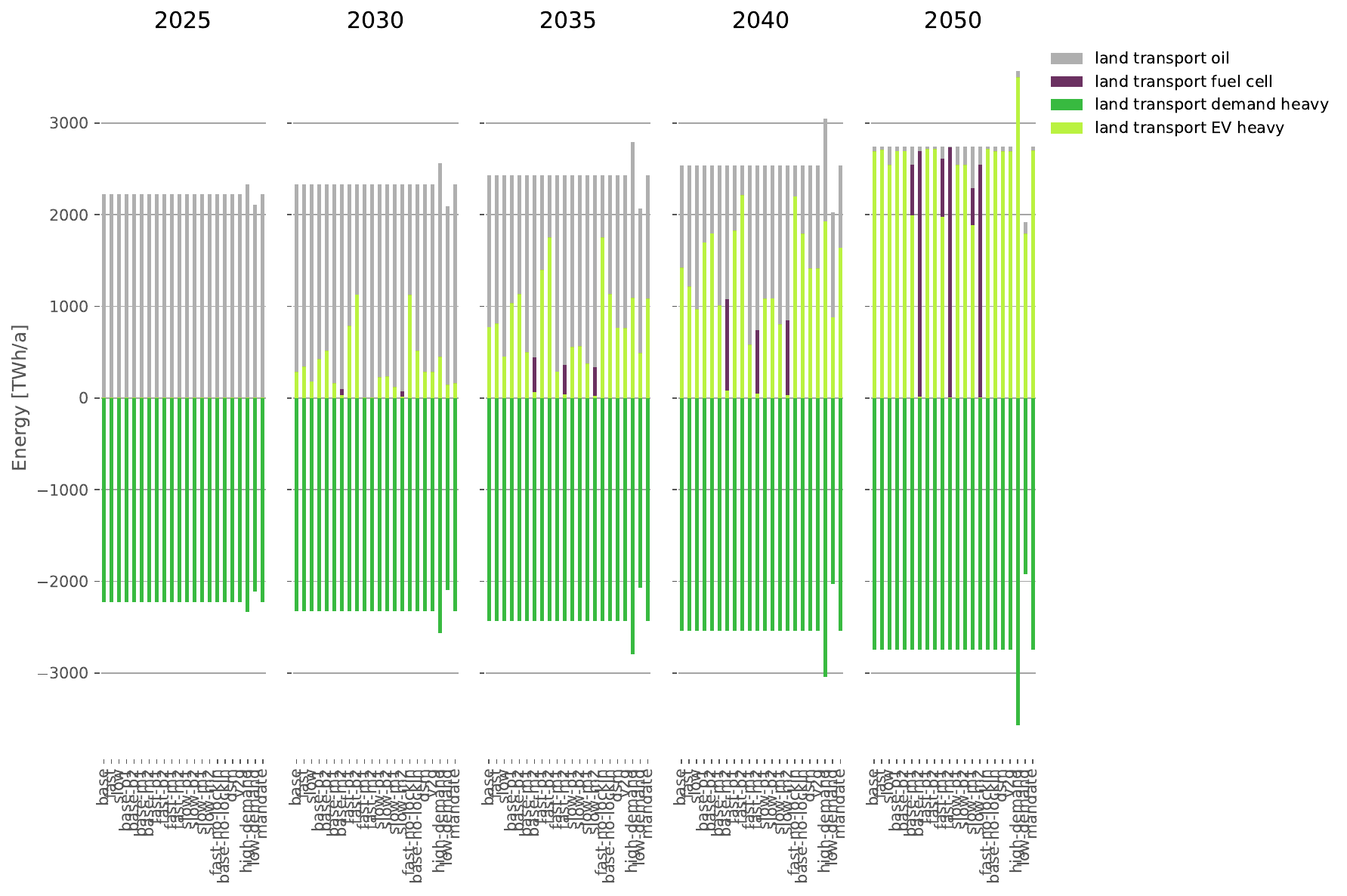}
		\caption{road transport heavy-duty}
		\label{fig:balances_heavy}
	\end{subfigure}
	\hfill
	\begin{subfigure}[b]{0.45\linewidth}
		\centering
		\includegraphics[width=\linewidth]{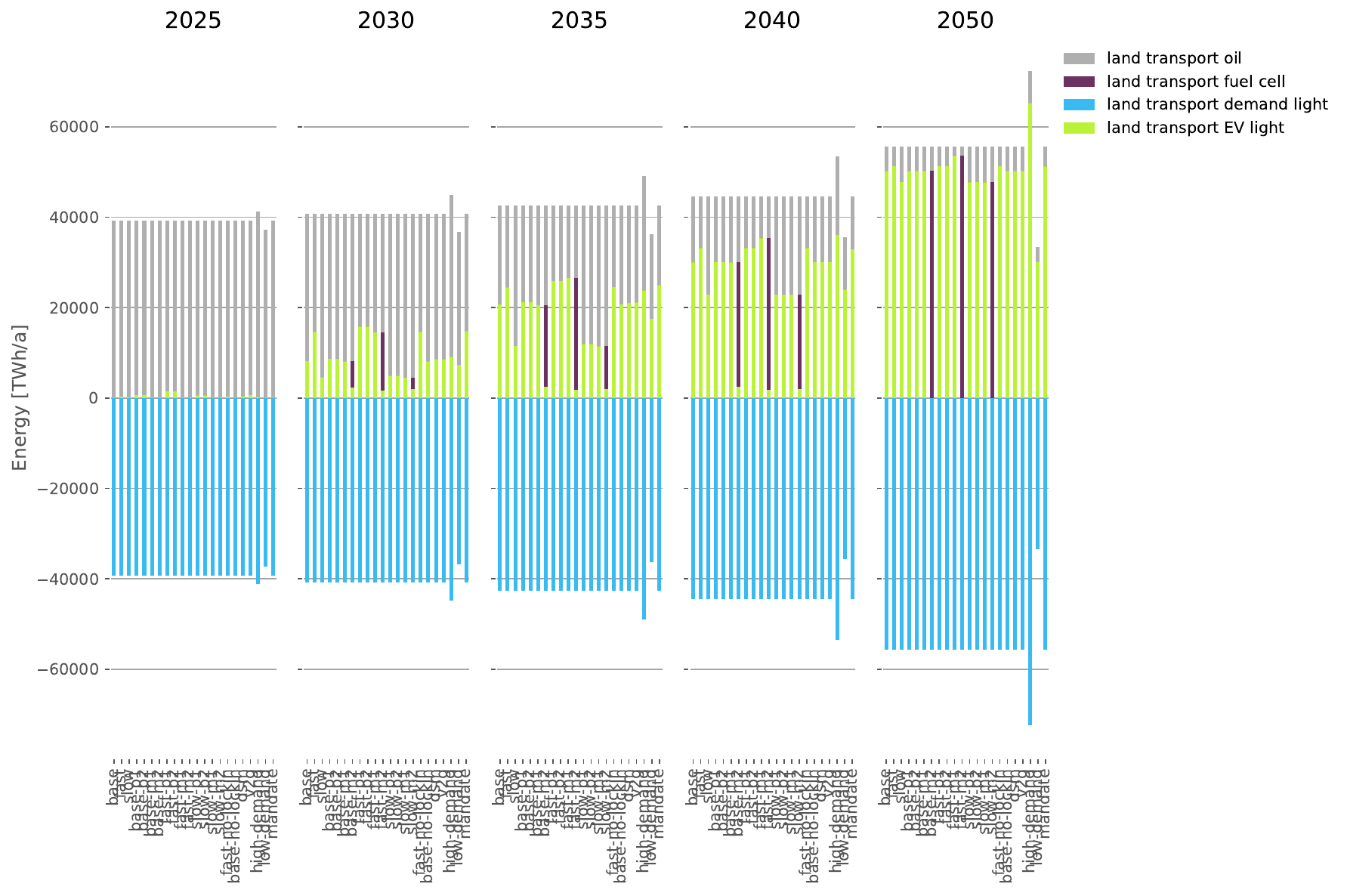}
		\caption{road transport light-duty}
		\label{fig:balances_light}
	\end{subfigure}
	\begin{subfigure}[b]{0.45\linewidth}
		\centering
		\includegraphics[width=\linewidth]{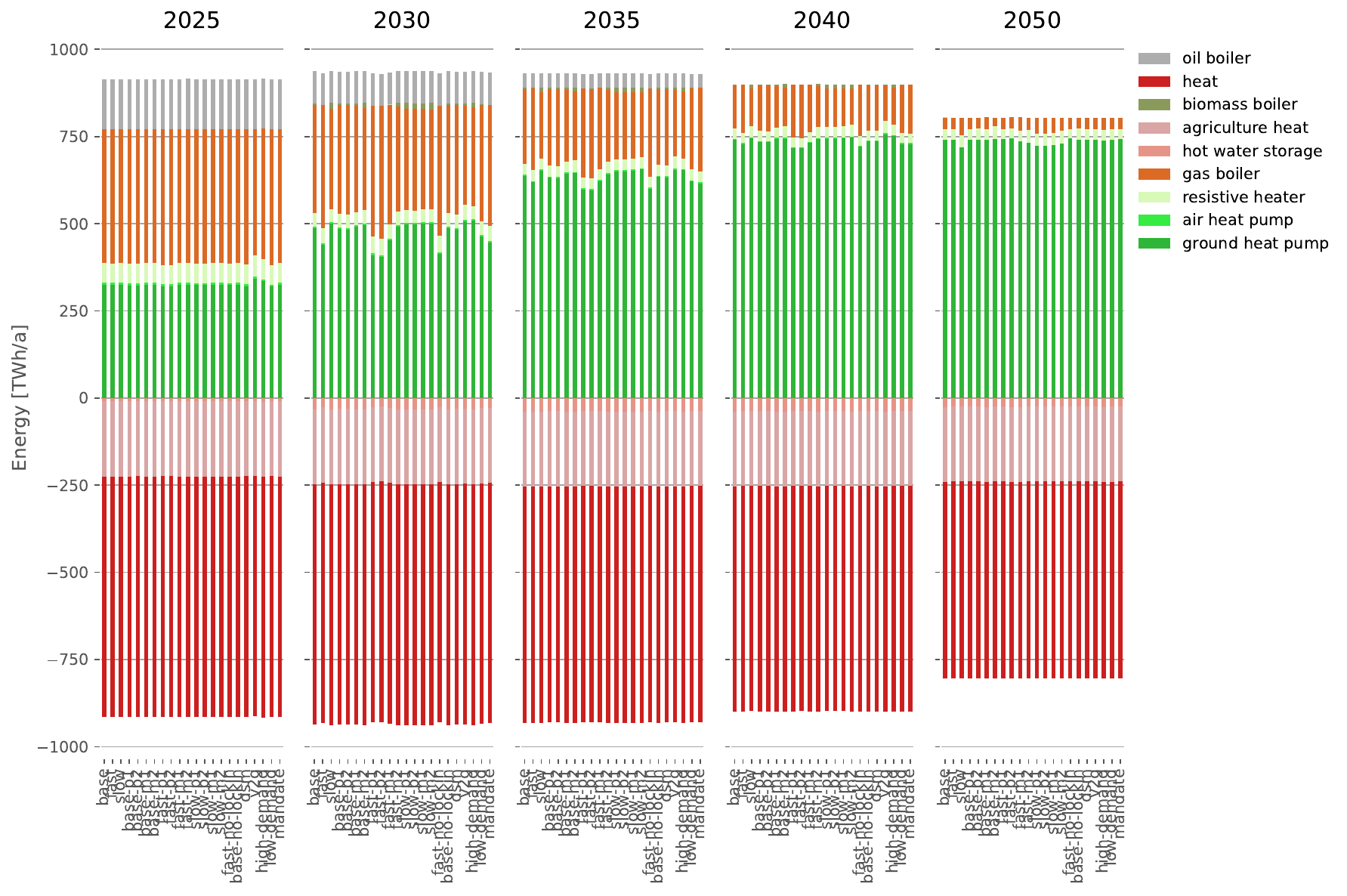}
		\caption{rural heat}
		\label{fig:balances_rural_heat}
	\end{subfigure}
	\hfill
	\begin{subfigure}[b]{0.45\linewidth}
		\centering
		\includegraphics[width=\linewidth]{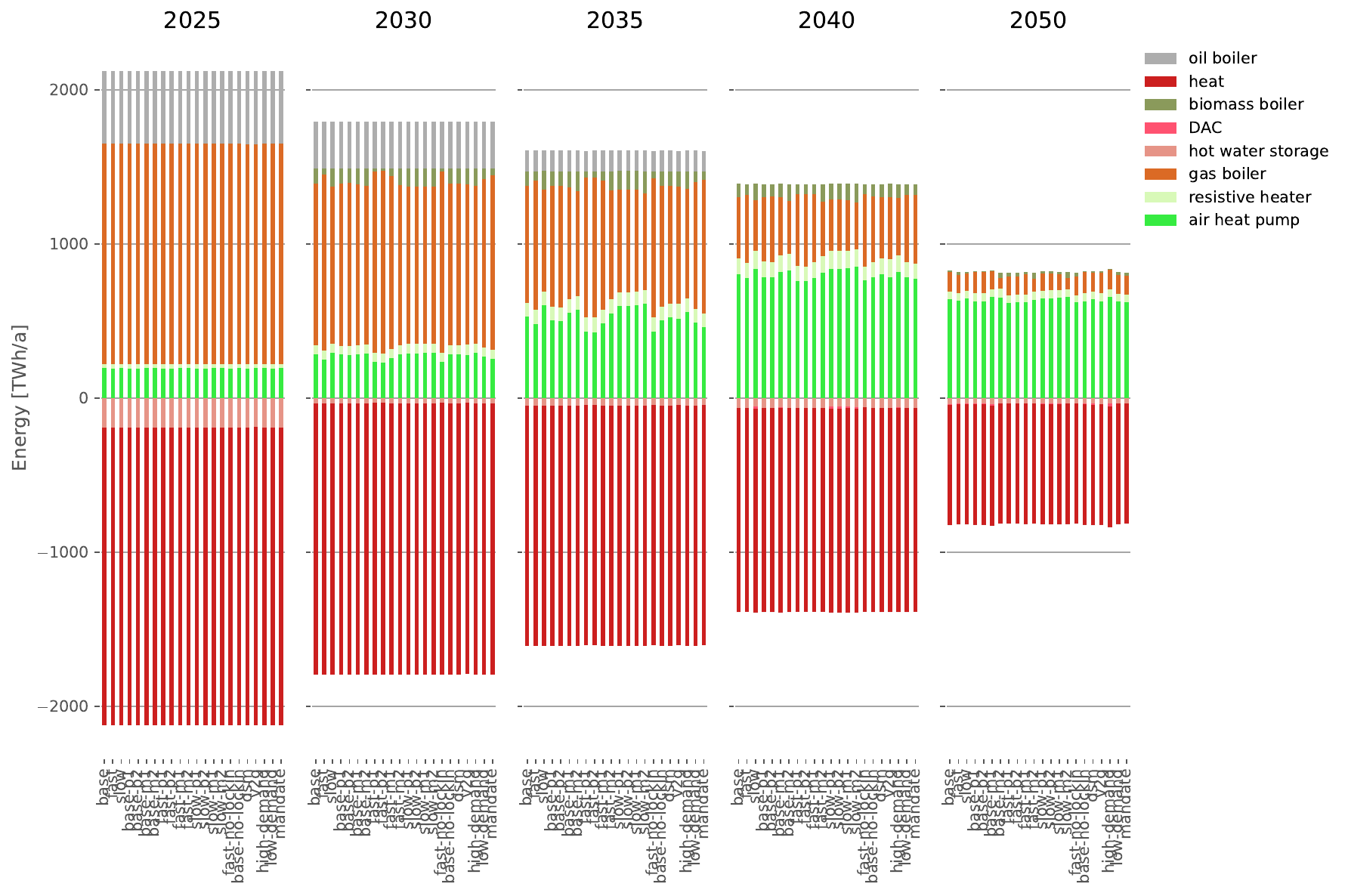}
		\caption{urban individual heating}
		\label{fig:balances_urban_decentral}
	\end{subfigure}
	\caption{Energy balances for various energy carriers.}
	\label{fig:balances}
\end{figure*}

\begin{figure}[h!]
	\centering
	\includegraphics[width=0.5\linewidth]{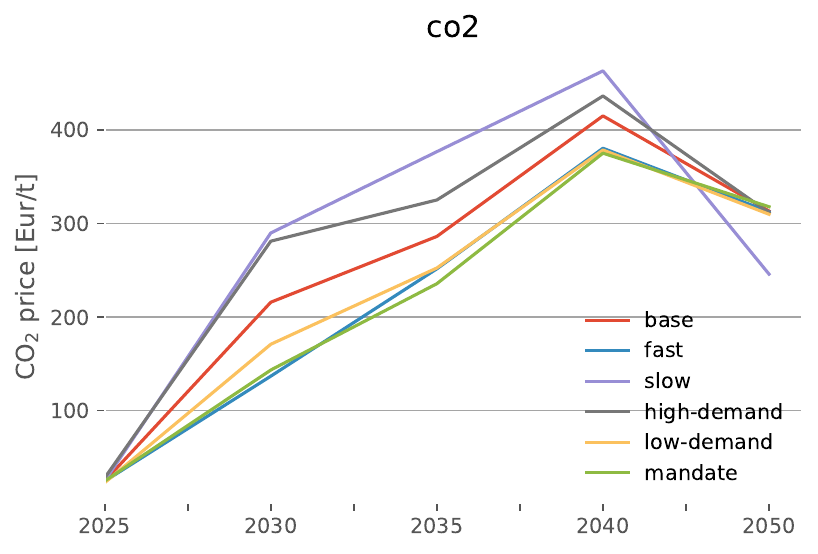}
	\caption{CO$_2$ price.}
	\label{fig:co2_price}
\end{figure}

\begin{figure*}[h!]
	\centering
	\begin{subfigure}[b]{0.45\linewidth}
		\centering
		\includegraphics[width=\linewidth]{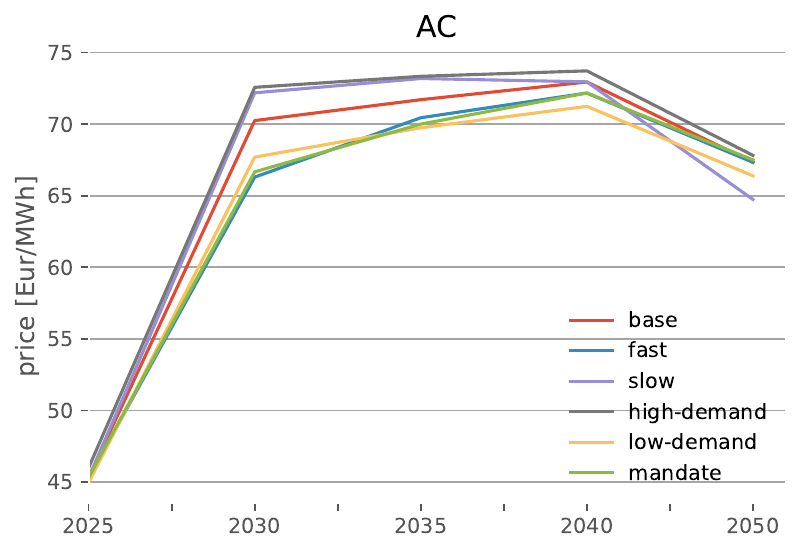}
		\caption{AC}
		\label{fig:prices_AC}
	\end{subfigure}
	\hfill
	\begin{subfigure}[b]{0.45\linewidth}
		\centering
		\includegraphics[width=\linewidth]{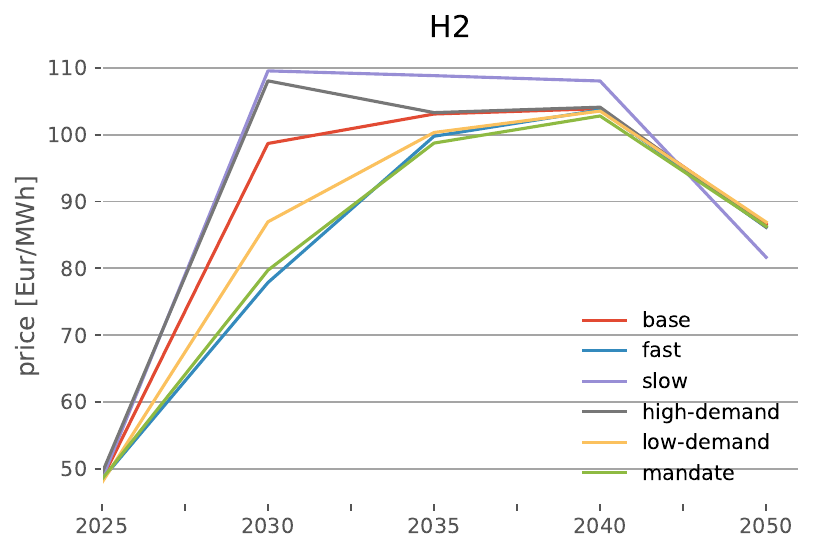}
		\caption{H$_2$}
		\label{fig:prices_H2}
	\end{subfigure}
\caption{Energy prices.}
\label{fig:prices}
\end{figure*}

The Appendix is structured as follows:
The Appendix is organized as follows:

\begin{enumerate}
	\item \textbf{Additional Graphics for Main Scenarios:}
	\begin{itemize}
		\item \textbf{Scenario Assumptions:}
		\begin{itemize}
			\item \textbf{Figure \ref{fig:scenariooverview}:} Overview of maximum allowed decarbonisation speeds and road transport demand.
			\item \textbf{Figure \ref{fig:max_share}:} Maximum share of zero-emission vehicles in newly-registered cars by scenario.
			\item \textbf{Figure \ref{fig:car_registration}:} Newly registered cars per country.
			\item \textbf{Figure \ref{fig:demand_historical}:} Historical demand growth in vehicle-kilometres driven for light- and heavy-duty vehicles.
			\item \textbf{Figure \ref{fig:transport_demand_example_week}:} Demand profiles for light- and heavy-duty vehicles.
			\item \textbf{Figure \ref{fig:transport_demand_total}:} Total transport demand per country.
			\item \textbf{Figure \ref{fig:heavy-duty-investmentcosts}} and \textbf{Table \ref{tab:vehicle_tech_assumptions}:} Technology assumptions for heavy-duty and light-duty vehicles.
		\end{itemize}
		\item \textbf{Main Results:}
		\begin{itemize}
			\item \textbf{Figures \ref{fig:total_costs} and \ref{fig:summed_costs}:} Total system costs split by technology and summed per investment period.
			\item \textbf{Figures \ref{fig:installation_scenarios} and \ref{fig:annual_additional_installation_scenarios}:} Installation of cumulative and additional renewable capacities and heat pumps.
			\item \textbf{Figure \ref{fig:co2-heatmap-demand}:} Comparison of CO$_2$ emissions between High-Demand and Low-Demand scenarios versus the Base scenario.
			\item \textbf{Figure \ref{fig:balances}:} Energy balances for various carriers (positive values represent supply, negative values represent demand).
			\item \textbf{Figure \ref{fig:co2_price} and Figure \ref{fig:prices}:} Energy prices for CO$_2$, electricity and hydrogen, resulting from the optimisation.
		\end{itemize}
	\end{itemize}
	
	\item \textbf{Sensitivity Analysis:}
	\begin{itemize}
		\item Section \ref{sec:sensi_v2g}: Analysis of \gls{v2g} and \gls{dsm} impacts.
	\end{itemize}
\end{enumerate}

\subsection{Scenario assumptions}

\begin{table}[]
	\begin{tabular}{llllllll}
		& \textbf{parameter} & \textbf{unit}         & \textbf{2025} & \textbf{2030} & \textbf{2035} & \textbf{2040} & \textbf{2050} \\
		\hline
		\multirow{5}{*}{\textbf{FCEV heavy}} & FOM                & \% of investment/year & 0.0           & 0.0           & 0.0           & 0.0           & 0.0           \\
		& efficiency         & kWh/km                & 2.16          & 2.05          & 1.94          & 1.83          & 1.61          \\
		& annualised         & EUR/vehicle/a         & 45741.15      & 33289.12      & 31489.08      & 29689.09      & 26089.08      \\
		& investment         & EUR/vehicle           & 368193.48     & 270752.69     & 255753.23     & 240753.76     & 210754.84     \\
		& lifetime           & years                 & 12.42         & 12.42         & 12.42         & 12.42         & 12.42         \\
		\hline
		\multirow{5}{*}{\textbf{ICE heavy}}  & FOM                & \% of investment/year & 0.0           & 0.0           & 0.0           & 0.0           & 0.0           \\
		& efficiency         & kWh/km                & 2.79          & 2.61          & 2.44          & 2.26          & 1.91          \\
		& annualised         & EUR/vehicle/a         & 21211.04      & 21211.04      & 21211.04      & 21211.04      & 21211.04      \\
		& investment         & EUR/vehicle           & 171404.74     & 171404.74     & 171404.74     & 171404.74     & 171404.74     \\
		& lifetime           & years                 & 12.42         & 12.42         & 12.42         & 12.42         & 12.42         \\
		\hline
		\multirow{5}{*}{\textbf{BEV heavy}}  & FOM                & \% of investment/year & 0.0           & 0.0           & 0.0           & 0.0           & 0.0           \\
		& efficiency         & kWh/km                & 1.17          & 1.1           & 1.04          & 0.97          & 0.83          \\
		& annualised         & EUR/vehicle/a         & 38862.64      & 27779.79      & 26657.66      & 25535.57      & 23291.32      \\
		& investment         & EUR/vehicle           & 313920.96     & 225373.29     & 215936.83     & 206500.37     & 187627.45     \\
		& lifetime           & years                 & 12.42         & 12.42         & 12.42         & 12.42         & 12.42         \\
		\hline \hline
		\multirow{5}{*}{\textbf{FCEV light}} & FOM                & \% of investment/year & 1.1           & 1.1           & 1.1           & 1.2           & 1.2           \\
		& efficiency         & kWh/km                & 0.33          & 0.33          & 0.33          & 0.33          & 0.33          \\
		& annualised         & EUR/vehicle/a         & 5254.57       & 4013.52       & 3710.81       & 3585.63       & 3273.84       \\
		& investment         & EUR/vehicle           & 43500.0       & 33226.0       & 30720.0       & 29440.0       & 26880.0       \\
		& lifetime           & years                 & 15.0          & 15.0          & 15.0          & 15.0          & 15.0      
		\\
		\hline
		\multirow{5}{*}{\textbf{ICE light}}  & FOM                & \% of investment/year & 1.6           & 1.6           & 1.6           & 1.6           & 1.6           \\
		& efficiency         & kWh/km                & 0.62          & 0.62          & 0.62          & 0.62          & 0.62          \\
		& annualised         & EUR/vehicle/a         & 3057.94       & 3144.74       & 3223.11       & 3291.67       & 3381.36       \\
		& investment         & EUR/vehicle           & 24309.0       & 24999.0       & 25622.0       & 26167.0       & 26880.0       \\
		& lifetime           & years                 & 15.0          & 15.0          & 15.0          & 15.0          & 15.0      \\ 
		\hline
		\multirow{5}{*}{\textbf{BEV light}}  & FOM                & \% of investment/year & 0.9           & 0.9           & 0.9           & 0.9           & 0.9           \\
		& efficiency         & kWh/km                & 0.19          & 0.19          & 0.19          & 0.19          & 0.19          \\
		& annualised         & EUR/vehicle/a         & 3422.71       & 2925.2        & 2893.6        & 2862.0        & 2798.92       \\
		& investment         & EUR/vehicle           & 28812.0       & 24624.0       & 24358.0       & 24092.0       & 23561.0       \\
		& lifetime           & years                 & 15.0          & 15.0          & 15.0          & 15.0          & 15.0          \\
	\end{tabular}
	\caption{Vehicle technology assumptions.}
	\label{tab:vehicle_tech_assumptions}
\end{table}
\FloatBarrier
\subsection{Further graphics of the main results}
\begin{figure*}[h!]
	\centering
	\begin{subfigure}[b]{0.45\linewidth}
		\centering
		\includegraphics[width=\linewidth]{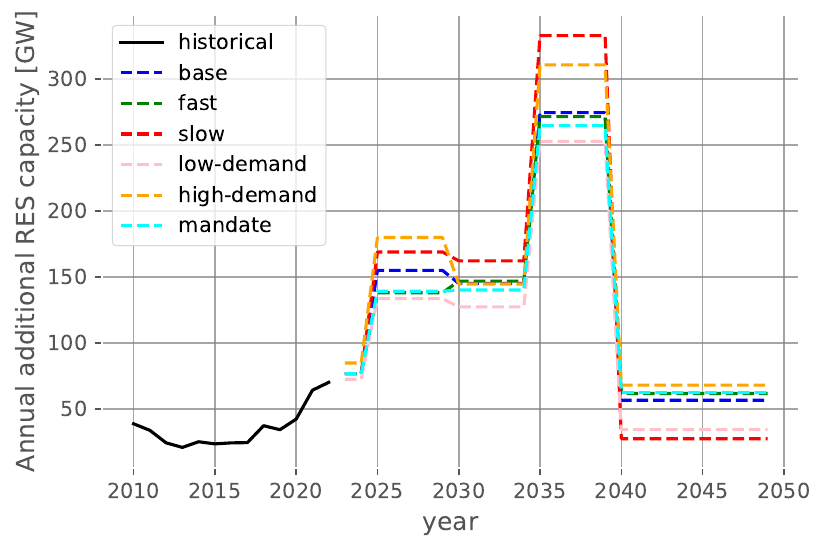}
		\caption{Annual additional renewable capacities.}
		\label{fig:annual_additional_res}
	\end{subfigure}
	\hfill
	\begin{subfigure}[b]{0.45\linewidth}
		\centering
		\includegraphics[width=\linewidth]{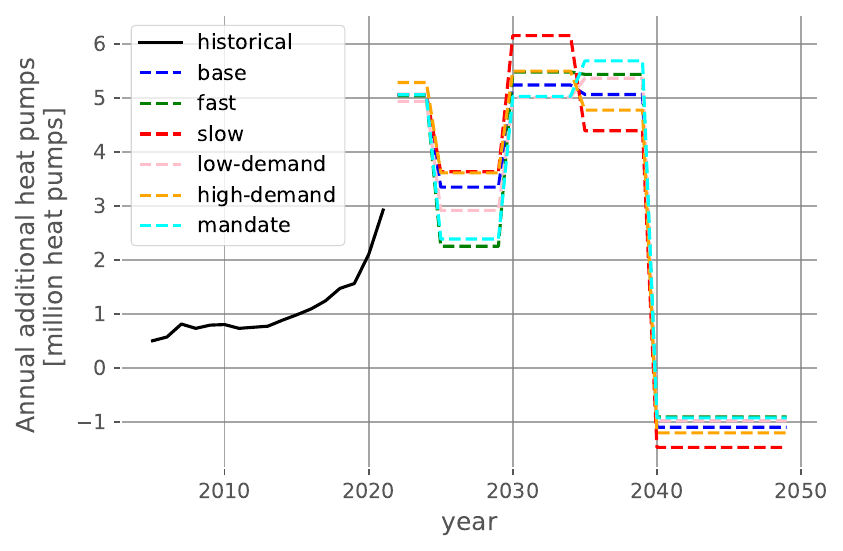}
		\caption{Annual additional number of heat pumps.}
		\label{fig:annual_additional_heat_pumps}
	\end{subfigure}
	
	\caption{Annual additional capacity depending on the scenario.}
	\label{fig:annual_additional_installation_scenarios}
\end{figure*}

\FloatBarrier
\FloatBarrier
\subsection*{Sensitivity analysis V2G and DSM}\label{sec:sensi_v2g}
In the main results, we made the conservative assumption that there is no \gls{dsm} of the \gls{bev}s and that they cannot charge back to the distribution grid (\gls{v2g}). In the following, we compare our \textbf{Base} scenario to cases where just \gls{dsm} or \gls{dsm} and \gls{v2g} are allowed. Flexible charging alone does not significantly impact the overall system design. At the same time, \gls{v2g} can reduce energy storage capacities of hydrogen and batteries, decrease the burden on the electricity distribution grid, and lower total annualized system costs marginally by up to -0.4\%. Over the modelling horizon, this leads to marginal cumulative savings of 9 and 130~billion \officialeuro for \gls{dsm} and \gls{v2g}, respectively (0.02\% and 0.3\% of total cumulative costs), assuming a social discount rate of 2\%.

\begin{figure}[h]
	\centering
	\includegraphics[width=0.7\linewidth]{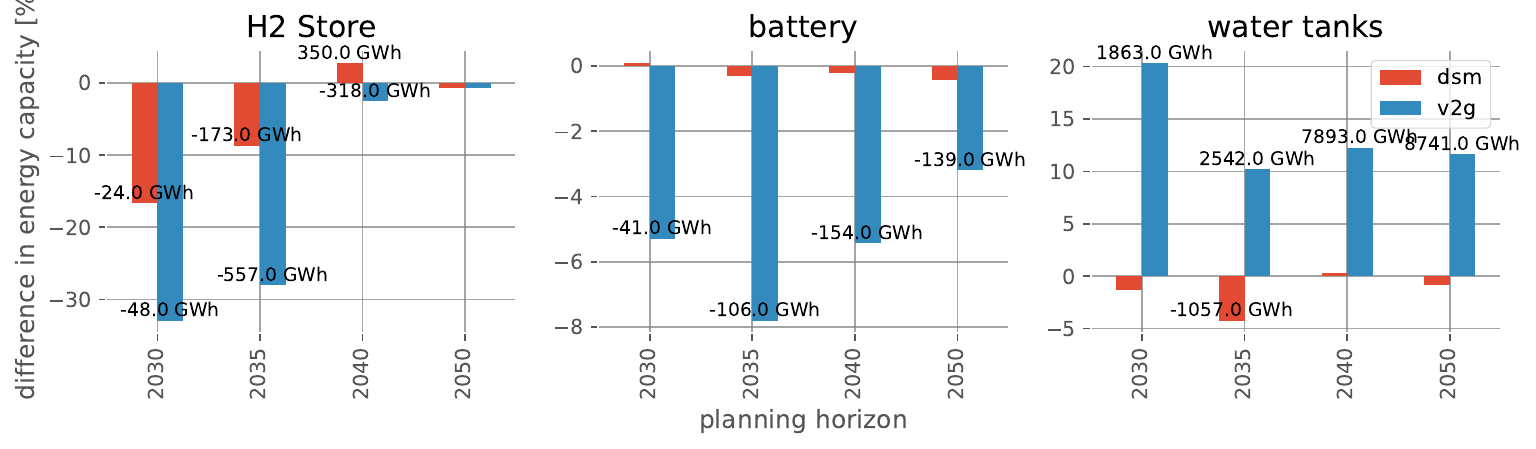}
	\caption{Changes in store energy capacities if \gls{dsm} or \gls{v2g} is available compared to the Base scenario.}
	\label{fig:storesdsmv2g}
\end{figure}

Flexible charging of the \gls{bev} allows lowering the energy storage capacity of the hydrogen storage in the near-term by more than 15\% and 30\% in case of \gls{dsm} and \gls{v2g} respectively. Further stationary battery storage can be reduced in scenarios that allow for \gls{v2g} by 3--8\% compared to the \textbf{Base} scenario. Thermal storage in the form of water tanks is used in these scenarios to a larger extent since the heat pumps are running more flexibly and the heat is stored (see Figure \ref{fig:storesdsmv2g}). 

\begin{figure}[h]
	\centering
	\begin{subfigure}[t]{0.5\linewidth}
		\centering
		\includegraphics[width=\linewidth]{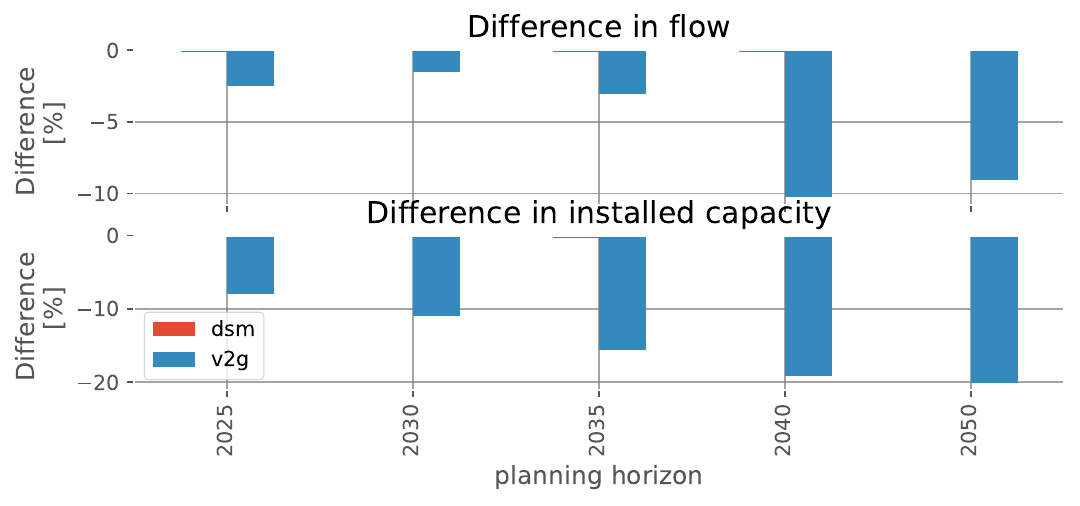}
		\caption{Changes in flow and installed capacity}
		\label{fig:elec-dist-diff-dsm-v2g}
	\end{subfigure}
	\hfill
	\begin{subfigure}[t]{0.4\linewidth}
		\centering
		\includegraphics[width=\linewidth]{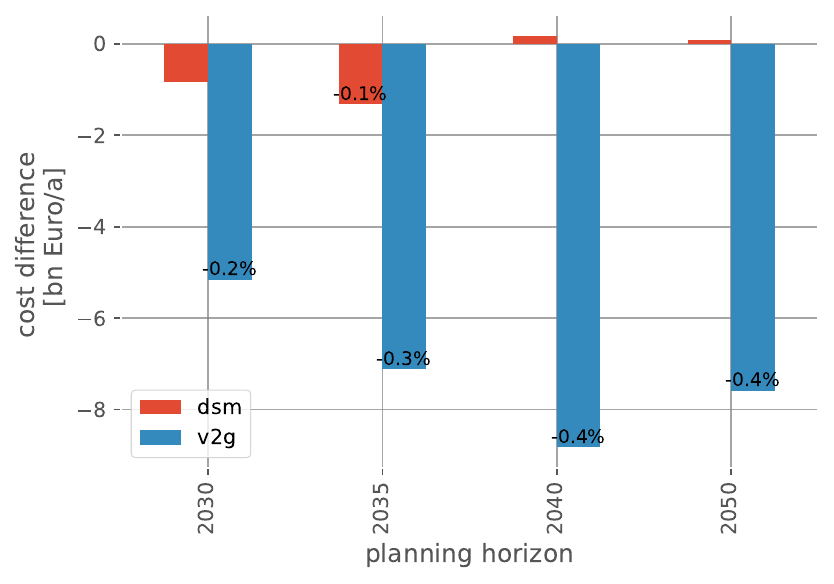}
		\caption{Changes in total system costs}
		\label{fig:cost-diff-dsm-v2g}
	\end{subfigure}
	\caption{Changes compared to the Base scenario.}
	\label{fig:comparison_dsn-v2g}
\end{figure}

If \gls{v2g} is available, the capacity of the electricity distribution grid can be reduced by up to 20\% since vehicles could be discharged at times of high demands within the building and charged at times with a higher feed-in of rooftop PV (see Figure \ref{fig:comparison_dsn-v2g}).


\end{document}